\documentclass[acmsmall]{acmart}

\AtBeginDocument{%
  }

\title{An Empirical Analysis of Machine Learning Model and Dataset Documentation, Supply Chain, and Licensing Challenges on Hugging Face}
\author{Trevor Stalnaker}
\orcid{0009-0005-6000-4227}
\affiliation{%
  \institution{William \& Mary}
  \city{Williamsburg, VA}
  \country{USA}
}
\email{twstalnaker@wm.edu}

\author{Nathan Wintersgill}
\orcid{0009-0006-2123-7412}
\affiliation{%
  \institution{William \& Mary}
  \city{Williamsburg, VA}
  \country{USA}
}
\email{njwintersgill@wm.edu}

\author{Oscar Chaparro}
\orcid{0000-0003-2838-685X}
\affiliation{%
  \institution{William \& Mary}
  \city{Williamsburg, VA}
  \country{USA}
}
\email{oscarch@wm.edu}

\author{Laura A. Heymann}
\orcid{0000-0001-9258-2105}
\affiliation{%
  \institution{William \& Mary}
  \city{Williamsburg, VA}
  \country{USA}
}
\email{laheym@wm.edu}

\author{Massimiliano Di Penta}
\orcid{0000-0002-0340-9747}
\affiliation{%
  \institution{University of Sannio}
  \city{}
  \country{Italy}
}
\email{dipenta@unisannio.it}

\author{Daniel M German}
\orcid{}
\affiliation{%
  \institution{University of Victoria}
  \city{Victoria}
  \country{Canada}
}
\email{dmg@uvic.ca}

\author{Denys Poshyvanyk}
\orcid{0000-0002-5626-7586}
\affiliation{%
  \institution{William \& Mary}
  \city{Williamsburg, VA}
  \country{USA}
}
\email{denys@cs.wm.edu}

\date{November 2024}

\usepackage{ifthen}

\newboolean{showcomments}
\setboolean{showcomments}{false}

\newboolean{showboxes}
\setboolean{showboxes}{true}

\usepackage{verbatim}

\usepackage[utf8]{inputenc}
\usepackage{url}
\usepackage{colortbl}
\usepackage{balance}
\usepackage{xspace}
\usepackage{graphicx}
\usepackage{bold-extra}
\usepackage{paralist}
\usepackage{multirow}
\usepackage{listings}
\usepackage{boxedminipage}
\usepackage{soul}
\usepackage{enumitem}
\usepackage[normalem]{ulem}
\setlist{nolistsep,leftmargin=.5cm}
\usepackage{booktabs}
\usepackage{tabularx}
\usepackage{svg}
\usepackage{calc}
\usepackage{float}
\usepackage{multirow}
\usepackage[normalem]{ulem}
\useunder{\uline}{\ul}{}
\usepackage[most]{tcolorbox}
\usepackage{caption}
\usepackage{microtype} 
\usepackage{algorithmic}
\usepackage{textcomp}
\usepackage{xcolor}

\usepackage{wrapfig}
\usepackage{subcaption}
\usepackage{cleveref} 
\usepackage{adjustbox}
\usepackage{totcount}

\usepackage{breakurl}

\usepackage{hyperref}

\ifthenelse{\boolean{showcomments}}
{\newcommand{\nb}[2]{
		\fbox{\bfseries\sffamily\scriptsize#1}
		{\sf\small$\blacktriangleright$\textit{#2}$\blacktriangleleft$}
	}
	
}
{\newcommand{\nb}[2]{}
	
}

\newcommand\add[1]{{\color{black}{#1}\xspace}}

\newcommand\nathan[1]{{\color{olive} \nb{NATHAN}{#1}}}
\newcommand\laura[1]{{\color{violet} \nb{LAURA}{#1}}}

\newcommand{\ie}{\textit{i.e.},\xspace}
\newcommand{\eg}{\textit{e.g.},\xspace}
\newcommand{\etc}{\textit{etc.}\xspace}
\newcommand{\etal}{\textit{et al.}\xspace}

\newcommand{\model}[1]{\textsc{{#1}}\xspace}
\newcommand{\license}[1]{\textsc{{#1}}\xspace}
\newcommand{\library}[1]{\textsc{{#1}}\xspace}
\newcommand{\user}[1]{\textit{{#1}}\xspace}
\newcommand{\cardData}{\textsc{CardData}\xspace}
\newcommand{\unknown}{``Unknown''\xspace}
\newcommand{\other}{``Other''\xspace}

\newcommand{\subsubsubsection}[1]{\underline{{#1}: }\xspace}

\newcommand{\supplychain}{ML supply chain\xspace}

\newcounter{findingcounter}
\newtotcounter{totalfindings}
\ifthenelse{\boolean{showboxes}}
{
    \newcommand{\finding}[1]{%
      \refstepcounter{findingcounter}
      \begin{tcolorbox}[boxsep=1pt,left=2pt,right=2pt,top=1pt,bottom=1pt]%
      \small
      \textbf{Finding \arabic{findingcounter}:} #1
      \end{tcolorbox}%
      \addtocounter{totalfindings}{1}
    }
    
  }
{
    \newcommand{\finding}[1]{}
}

\begin{abstract}

The last decade has seen widespread adoption of Machine Learning (ML)
components in software systems. This has occurred in nearly every domain, from natural language processing to computer vision. These ML components range from relatively simple neural networks to complex and resource-intensive large language models.  However, despite this widespread adoption, little is known about the supply chain relationships that produce these models, which can have implications for compliance and security.  In this work, we conducted an extensive analysis of 760,460 models and 175,000 datasets extracted from the popular model-sharing site Hugging Face. First, we evaluate the current state of documentation in the Hugging Face supply chain, report real-world examples of shortcomings, and offer actionable suggestions for improvement.  Next, we analyze the underlying structure of the existing supply chain.  Finally, we explore the current licensing landscape against what was reported in previous work and discuss the unique challenges posed in this domain.  Our results motivate multiple research avenues, including the need for better license management for ML models/datasets, better support for model documentation, and automated inconsistency checking and validation.  We make our research infrastructure and dataset available to facilitate future research.

\end{abstract}

\begin{document}

\maketitle

\section{Introduction}
\label{sec:intro}

The use of machine learning (ML) models in software applications has increased dramatically over the last decade, including in recommendation systems, computer vision, chatbots, image generation, and automated software engineering. As creating and training new models has become increasingly more expensive~\cite{patterson_carbon_2021}, developers have turned to fine-tuning pre-existing models.  This approach may save time and effort, but it also introduces the complexity of a new ML supply chain, complete with many novel challenges.%

Supply chains for conventional software typically consist of %
software components, libraries, configuration files, and processes that contribute to the software~\cite{williams2025research, okafor2022sok, wangmanaging}. %
However, for ML-intensive systems, the typical ML supply chain is more complex, comprising all elements that make an ML algorithm work~\cite{tan2022exploratory, GaoHXZ24}. Creating a modern ML system involves the integration of ML models and traditional software components, %
taking into account dependencies between models themselves or between models and their training datasets. In addition, the training of ML models can rely on multiple datasets or dataset aggregates, and datasets, in turn, each have their own supply chain. Because proper license compliance requires developers to consider the permissions granted and requirements imposed by any individual component's license, the complexity of ML supply chains raises new and challenging issues, particularly given that many questions remain regarding the applicability of copyright law to generative AI~\cite{lee2023talkin}.%

Adaptation or reuse of existing models can also introduce degrees of complexity~\cite{davis2023reusing}. Models can be fine-tuned~\cite{church2021emerging} or quantized from existing models~\cite{gholami2022survey}. %
(Fine-tuning allows a developer to use an existing model to create a model that specializes in certain tasks or to change the model's behavior, such as removing safety or ethical protections. Quantization compresses an existing model such that it can be run on smaller hardware with minimal performance loss.)  Different ML models can be combined to form a single architecture that can, in turn, be reused or adapted~\cite{yang2024model}. Finally, the outputs from one model can be used to train another model, such as through synthetic data~\cite{marwala2023use} or the use of distillation~\cite{hinton2015distilling}. (Distillation is a technique where a smaller, simpler student model is trained to mimic the behavior of a larger, more complex teacher model.)

The ML supply chain, particularly those involving generative models \cite{lee2023talkin}, is not necessarily linear, progressing cleanly from one dependency or stage to the next.  Instead, as outlined by Lee \etal  \cite{lee2023talkin}, there can be branches and even cycles in the supply chain. %
For example, datasets can be used not only in the initial training of models but also for the fine-tuning of pre-existing base models, sometimes in multiple instances and by different developers. Additionally, some stages of the supply chain can backfeed into others, making relationships increasingly complex, if not recursive.  For example, the outputs of a generative model can be added to pre-existing datasets and used for training future versions.  The sheer volume of content from generative models makes this situation increasingly likely, if not inevitable. 

Understanding these relationships is critical not only for license compliance, which requires a full understanding of the components used in a project and their associated licenses, but also for detecting, mitigating, and managing security threats involved with model reuse, such as weight poisoning attacks \cite{Kurita2020WeightPA}, data poisoning \cite{Goldblum2020DatasetSF}, and even malware hidden in model weights and assembled at runtime \cite{Wang2021EvilModelHM}.

 While previous work has considered the documentation \cite{pepe2024hugging}, evolution \cite{jiang2022empirical}, and environmental impact \cite{castano2024lessons} of ML models, the complexity and challenges of the ML supply chain as a whole have not yet been fully explored. The supply chain relationships between models and dependent GitHub repositories have been explored by Pepe \etal \cite{pepe2024hugging} and Jiang \etal \cite{jiang2024peatmoss}, %
 but the relationships between ML models remain understudied.

To contribute to bridging this gap, we aim to investigate the emerging ML model supply chain on Hugging Face~\cite{hugging-face} (hereafter referred to as the ``ML supply chain''%
), specifically the documentation practices of model owners, the structure and complexity of the supply chain itself, and the existence and prevalence of potential license compliance issues. Launched in 2016, Hugging Face is, as of September 2025, %
the largest repository of ML models and datasets~\cite{Jiang2023AnES, jiang2022empirical}. As with GitHub and traditional software, Hugging Face provides a central hub for model developers and data scientists to explore, share, and experiment with ML models.  While other model forges---such as TensorFlow Hub~\cite{tensorflowHub}, PyTorch Hub~\cite{pytorchHub}, and Model Zoo~\cite{modelzoo}---exist, the Hugging Face platform has the widest reach, hosting more than 750K models and 175K datasets across many different use cases. Hugging Face therefore represents a robust opportunity to understand ML supply chains and analyze how developers interact with them.

We mine models and datasets hosted on Hugging Face, using the available information to construct a supply chain graph.  %
We highlight different types of documentation inconsistencies that complicate the creation of this supply chain graph, discuss the underlying structure and features of the current ML supply chain on Hugging Face, and analyze the present licensing landscape. We note that proper documentation and documentation practices are foundational to mapping, and thus managing, the ML supply chain. Ultimately, we demonstrate that there is still much work to be done to facilitate better dependency/supply chain management on Hugging Face and in the ML ecosystem.%

The main contributions of this paper are the following:
\begin{itemize}
    \item A critical evaluation of the current state of documentation on Hugging Face.
    \item An analysis of the interactions between models and datasets in the ML supply chain on Hugging Face.
    \item An analysis of the licensing ecosystem on Hugging Face and potential compliance problems.
    \item An in-depth manual analysis of the top models and datasets on Hugging Face. %
    \item Actionable suggestions for how model hubs can better facilitate supply chain management activities.
    \item A cleaned, reusable, and extensible dataset of models, datasets, and their relationships, and a tool suite capable of generating newer versions of said dataset~\cite{anonymous_repo}.
\end{itemize}

\add{\section{Background}
\label{sec:back}
In this section, we provide the background necessary to contextualize and understand our results. Specifically, we highlight the ML supply chain, software licensing, and documentation practices for ML components on Hugging Face.

\subsection{The ML supply chain}

The ML supply chain has become a complex structure with many steps and components, each of which can affect the final output of a given model. Lee \etal~\cite{lee2023talkin} identify eight events in the generative AI supply chain, but most apply to AI systems more broadly%
: 
\begin{enumerate}
\item The creation of expressive works that will eventually be used to train a model;
\item The conversion of these expressive works into digital data;
\item The compilation of these data points into training datasets;
\item The creation of an ML model by selecting an architecture, training datasets, and a training algorithm;
\item The fine-tuning of existing ``base'' models;
\item The deployment of the trained/fine-tuned model as a service;
\item The use of the model to generate output, and finally 
\item Applying additional alignment to further improve the model or meet user needs.
\end{enumerate}

More generally, the ML supply chain comprises various components, including datasets, models, and traditional software.  These components relate to each other in several key ways, including data examples being collected to create a dataset, multiple datasets being aggregated to create a composite dataset, datasets being used to train ML models, models being fine-tuned on pre-trained base models, and models being incorporated into software projects. %
These complex relationships distinguish the ML supply chain from other software supply chains: different steps in the process of creating and deploying ML models can feed back into each other in various, nonlinear ways, as identified by Lee \etal~\cite{lee2023talkin}, and different types of dependency relationships can relate to or build off of the original works in different ways, as identified by Duan \etal~\cite{duan2024modelgo}.
Understanding how work relates is important, as reusing any component through a supply chain can raise security risks or licensing implications~\cite{wangmanaging}.

Modern ML models are also becoming increasingly complex with respect to their architectures.  For example, Mixture of Expert (MoE) models and ensemble models comprise multiple specialized models trained by dividing the problem space into homogeneous regions~\cite{masoudnia2014mixture, mixtral}.  Recent work has also proposed using smaller ML models to replace the hidden states within a Recurrent Neural Network (RNN) architecture to facilitate better recall and compression of context \cite{sun2024learning}.  Both examples demonstrate that understanding the relationships among models and model architectures will become increasingly difficult.

Hugging Face, an online platform that hosts ML components, including models and datasets, has become a central hub for those who use and create ML components, showing exponential growth in recent years~\cite{castano2024analyzing}.   Users can download components provided by others and upload their own components. It has become the leading registry for pre-trained models online~\cite{Jones2024WhatDW}, presenting new opportunities for reusing ML components~\cite{davis2023reusing}. By studying the Hugging Face environment, we aim to uncover how multiple ML components can relate to one another and consider the implications of this critical supply chain.%

\subsection{Software Licensing}
In the same way as other creative works, software is protected by copyright law in the United States~\cite{copyright_software}.%
A copyright holder has several rights under U.S. copyright law, including the right to reproduce the work, create derivative works, distribute the work, and authorize others to engage in these activities~\cite{USCode17sect107}. Although the copyright owner can transfer some or all of the rights in a work, an owner who wishes to retain copyright while authorizing use by another party in a way that would otherwise be infringing typically does so by means of a license, which allows the copyright owner to impose conditions on the use~\cite{meeker2017open}. 
In practice, developers often select pre-existing licenses drafted by other organizations and entities rather than drafting new licenses~\cite{almeida2019investigating, wintersgill2024law}. This means that a particular license may or may not be a good fit for the subject of the license, the intended uses, or the copyright status of the underlying work.\laura{Added this last sentence to respond to Trevor's early query about whether we should acknowledge the uncertain status of, e.g., databases, without making things too complicated. Sound OK?} \nathan{Looks great, thank you} %

The licensing landscape for ML models, particularly in Hugging Face, is very diverse, including open source licenses (such as \license{GPL} and \license{MIT}) and Creative Commons licenses, as well as new ML-specific licenses such as \license{Open-RAIL} and \license{Llama}. %
These licenses broadly fall into two categories: permissive and restrictive  (sometimes called copyleft). Permissive licenses (such as \license{MIT}~\cite{mit},  \license{Apache-2.0}~\cite{apache}, and \license{CC-BY-4.0}~\cite{cc-by-4.0}) typically impose few restrictions, often only requiring the user to provide notice files containing attribution or license information.  On the other hand, copyleft licenses (such as GPL~\cite{gpl_license} and \license{CC-BY-SA-4.0}~\cite{cc-by-sa-4.0}) typically require that derivative works be made available under the same or compatible licensing (leading some to characterize such licenses as ``viral'').  

ML-specific licenses follow a similar pattern, but often also impose additional restrictions on model usage, typically involving ethical issues, limiting commercial use beyond a certain number of active users, and prohibitions on using model output to train competing models~\cite{MLEthics,restrictedUsage}. These additional restrictions have led some to consider many ML-specific licenses as falling outside the open-source definition \cite{model_restrictions}. In response, more new licenses---such as OpenMDW \cite{openMDW,linuxOpenMDW}---have emerged, implementing the Model Openness Framework \cite{white2024modelopennessframeworkpromoting}, which aims to align model development with open-source principles.

License compliance is a task with which developers often struggle \cite{vendome2018distribute}. These challenges are likely heightened in the ML supply chain, given the uncertainties and ambiguities involved in interpreting OSS and CC licenses in the new ML context \cite{duan2024modelgo,ossForML}. Developers frequently treat license compliance as a lower priority, partly due to limited enforcement~\cite{wintersgill2024law}. However, non-compliance can lead to financial~\cite{cost1, cost2}, legal~\cite{vizio_1, author_lawsuit, andersenVsStability22,authorsVsChatGPT23}, and reputational~\cite{johndeere, gunningham2004social} consequences for developers and organizations, particularly organizations of sufficient size and prominence to attract attention for non-compliance. More specifically, to the extent that a licensed work is protected by copyright law, use beyond the scope of the license may constitute copyright infringement, which can result in litigation by the licensor, demands for payment or other remedies, and/or undesirable publicity that affects the licensee's standing in the community. \laura{Added one additional sentence to make clear to R3-type readers why this matters and the connection between licensing and copyright.}

\subsection{ML component documentation}
Comprehensive component documentation is essential to understanding the ML supply chain and maintaining license compliance. This need has become increasingly critical as the complexity of the components and the processes involved in their creation and deployment continue to grow. However, ML components do not always disclose data sources, making it difficult to comply with or keep track of all licensing obligations associated with each component~\cite{hassan2024rethinking}.  This can create conditions that lead to legal disputes over the use of copyrighted material in training ML models~\cite{ai_copyright}, including not only the use of copyrighted works for training without permission but also the use of the generations of one model to train another model, an activity that may not have been permissible under that model's licensing terms \cite{llama3_license}.  

Model cards \cite{mitchell2019model} have become the standard method for sharing information about ML models hosted on Hugging Face \cite{what_are_model_cards}, including intended uses, limitations, and datasets used in training  \cite{model_card_guidebook}. Although tools are in development to streamline the process \cite{model_card_creation_tool}, model card creation on Hugging Face is primarily a manual process in which a user enters documentation about a model into a predefined markdown template \cite{model_card_template}. This template is robust, including fields for model description, uses, bias, limitations, testing, and optional fields for citation, technical specifications, and environmental impact. However, the manual nature of documentation introduces a significant potential for human error, ambiguity, and incompleteness \cite{jiang2024peatmoss, jiang2023ptmtorrent}. In fact, the quality and adoption of model cards remain low \cite{bhat2023aspirations}, despite user studies and other efforts by Hugging Face \cite{model_card_user_studies} to improve the situation, and despite researchers proposing tools such as DocML \cite{bhat2023aspirations} to create and evolve model cards.
}

\add{\section{Related work}
\label{sec:related_work}

Hugging Face has been studied in recent years to investigate the maintenance and evolution of ML models~\cite{castano2024analyzing}, security risks in the ML supply chain \cite{jiang2022empirical}, the carbon footprint of ML models~\cite{castano2023exploring}, and the documentation of ML datasets and models \cite{pepe2024hugging,oreamuno2024state,longpre2023data}. 

\subsection{Model and dataset documentation}

Previous research has identified several documentation shortcomings on Hugging Face. Works by Oreamuno \etal and Jiang \etal highlight that the majority of models and datasets hosted on Hugging Face lack documentation and that less-popular components are more likely to have incomplete documentation~\cite{oreamuno2024state, Jiang2023AnES}. Additional studies have explored the fields most likely to be missing from the dataset documentation~\cite{yang2024navigating}, the presence and quality of the license information~\cite{longpre2023data, pepe2024hugging}, and the documentation of known model biases~\cite{pepe2024hugging}. %
In addition, some studies have considered other deficiencies, such as poorly chosen or erroneous model names (\eg nonadherence to conventional naming schemes that lead to confusion)~\cite{jiang2023exploring,jiang2024naming}.

We build on these previous analyses by identifying \textit{specific} and \textit{previously unreported} documentation failures, such as component reference ambiguities, encountered in our analysis of the ML supply chain on Hugging Face (\Cref{sec:rq0}). Our detailed analysis and enumeration of specific, concrete examples of incomplete and erroneous documentation go beyond previous work. These real-world observations can help model owners avoid common documentation pitfalls, empower model hubs to implement automated checks for common errors, and make researchers aware of potential noise in data mined from Hugging Face.

\subsection{The ML supply chain}

By analyzing discussion forums and a mining study of Hugging Face, Taraghi \etal \cite{taraghi2024deep} identified various challenges and benefits associated with model reuse. However, they did not explicitly examine the dependency relationships that result from reuse. Other studies, such as those by Pepe \etal \cite{pepe2024hugging} and Jiang \etal \cite{jiang2024peatmoss}, have primarily focused on the relationship between models and downstream GitHub repositories that utilize those models. In contrast, our work instead focuses on the relationships among pre-trained/base models,%
dependent models, and datasets. While Jones \etal do consider model-to-model relationships, they stop short of constructing a comprehensive supply chain graph~\cite{Jones2024WhatDW}. We extend this line of research by constructing such a graph and conducting a robust statistical analysis of its structure. We also conduct a detailed manual review of the most popular models (\Cref{ssec:case_study_models}) and datasets hosted on Hugging Face (\Cref{ssec:case_study_datasets}).

\subsection{Software Licensing}

Much previous work has investigated software licensing for traditional software, including software licensing issues~\cite{vendome2018distribute, kapitsaki2020developers, wu2017analysis} and practitioners' understanding of software licensing~\cite{almeida2019investigating, wintersgill2024law}. More recently, some work has investigated the licensing issues that arise when ML components are used by open-source projects. Works by Pepe \etal \cite{pepe2024hugging} and Jiang \etal \cite{jiang2024peatmoss} explore model licensing on Hugging Face, primarily reporting the presence of licensing information as well as the usage frequencies for those licenses. In the case of Jiang \etal \cite{jiang2024peatmoss}, this was done as a short demonstration illustrating the utility of their PEATMOSS dataset. Both works also consider the interplay of model licenses with the licenses of dependent GitHub repositories. Neither work, however, evaluates license incompatibilities between models or between models and datasets, as we do in this work. %
We extend these prior works by collecting and analyzing more recent data, which reflects the current state of model/dataset licensing.  

The work of Duan \etal \cite{duan2024modelgo} proposes a tool (ModelGo) and framework for identifying potential license compliance issues involving ML components but provides only artificial examples and offers no empirical analysis of the prevalence of the potential problems in practice. Instead, we consider real-world examples of potential license conflicts and offer a rich, %
empirical analysis, including a report of license usage statistics, an examination of the frequencies of license differences between parent and children models, and thorough case-studies on a sample of the top models and datasets from Hugging Face. We further expand on prior work by considering various license classes, multi-licensing, and failing to comply with license-imposed naming requirements. 

\subsection{Summary}
In conclusion, while our study builds on the basis of prior work \cite{castano2024lessons,yang2024navigating,jiang2024naming,jiang2023ptmtorrent,oreamuno2024state,Jiang2023AnES,pepe2024hugging,taraghi2024deep,jiang2023peatmoss}, our investigation focuses on the structure and characteristics of the ML supply chain itself and the challenges encountered in managing it. More specifically, while previous work examined deficiencies in documentation and licensing at the level of individual models, we investigate the ML supply chain at a higher level by examining how it is structured and how models relate. Additionally, where previous work has examined the relationships between projects and models, we examine the relationships among multiple models. Thus, our work is beneficial not only to software developers, as is the case for prior work, but also to data scientists, ML engineers, %
and anyone who wants to reuse models with complex dependencies. %
By providing an ecosystem-level examination of the ML supply chain, we provide these groups with the information necessary to understand how models relate to one another and evolve, licensing concerns regarding ML components on Hugging Face, and potential challenges one can face regarding model documentation.
}

\section{Study Methodology}
\label{sec:design}
\label{sec:methodology}

The \emph{goal} of this study was to evaluate the current state of documentation, structure, and licensing of the ML supply chain. The study \emph{context} consisted of 760,460 models and 175,000 datasets mined from Hugging Face. We aim to address the following research questions (RQs):

\begin{enumerate}[label=\textbf{RQ$_\arabic*$:}, ref=\textbf{RQ$_\arabic*$}, wide, labelindent=10pt,leftmargin=10pt, start=0]\setlength{\itemsep}{0.2em} 

    \item \label{rq:2}{\textit{What documentation deficiencies exist in Hugging Face model and data cards that potentially complicate mapping the ML supply chain?}} %
    To understand the ML supply chain on Hugging Face, we needed to collect and analyze Hugging Face metadata. Using these data, we identified several issues and challenges related to documentation that introduced hurdles to obtaining a comprehensive understanding of the ecosystem. We examine these challenges in detail because with poorly and inconsistently described models and datasets, developers may struggle to leverage the ML supply chain effectively. This preliminary research question discusses examples of noisiness issues occurring in the Hugging Face model and the data set documentation.
  
    \item \label{rq:1}{\textit{What is the structure of the ML supply chain?}} The ML supply chain, like any other supply chain, is characterized by dependencies across components. This research question looks at the complexities of such a dependency graph.%
  
    \item \label{rq:3}{\textit{What is the current licensing landscape for models and datasets and what potential compliance challenges does it pose?}}  ML models are software, so software licenses govern their use and redistribution. %
    As such, these models present new challenges in the area of license compliance. To understand these challenges, we examine the state of licensing in the ML supply chain on Hugging Face by 1) analyzing common licenses and license categories, 2) exploring licensing decisions made in the supply chain, %
    and 3) investigating potential compliance challenges, such as incompatibilities between parent/child model licenses, ambiguities introduced by undeclared, unknown, or custom licenses, and uncertainty surrounding how license terms apply in context.  %
\end{enumerate}

\begin{figure}[ht!]
\centering
\includegraphics[width=\linewidth]{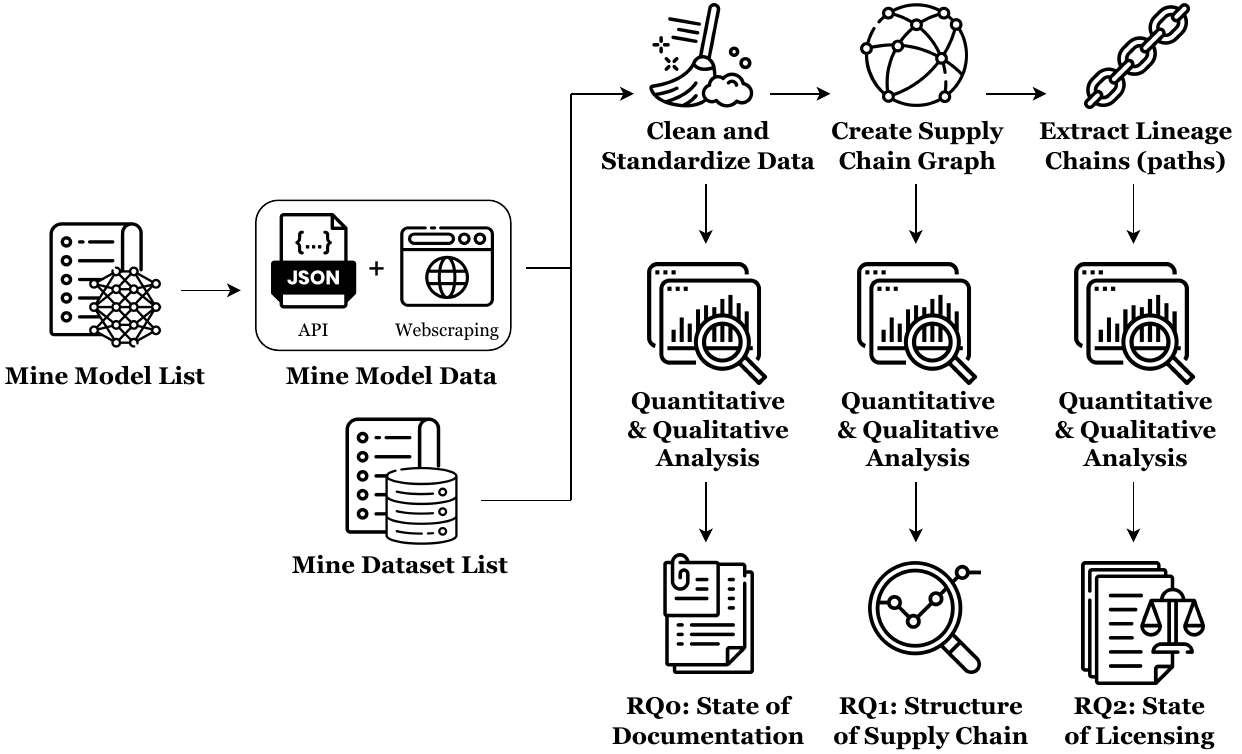}
\caption{Study Methodology} %
\label{fig:methodology}
\end{figure}

In the following subsections, we describe the study methodology, which is depicted in Fig. \ref{fig:methodology}. We start by mining model and dataset lists.  Next, for each model/dataset, we extract information necessary to construct the ML supply chain graph, recording and analyzing documentation deficiencies encountered during the process ($RQ_0$). Then we systematically analyze and evaluate the structure of the resulting supply chain graph ($RQ_1$).  %
The same graph, along with the licensing information, is then used to address $RQ_2$.

\subsection{Data collection}
\label{ssec:data_collection}
To answer the research questions, we used the Hugging Face API to extract a list of all models publicly hosted on the platform.  An initial list of 500,000 models was pulled on July 9, 2024, and a full list of all available models was obtained on July 11, 2024.  Using the same APIs, we mined data (model cards and metadata) for each model on this list.  For models that were unavailable through the API (as described in \Cref{sec:no_api_access}), we employed a web-scraping technique relying on a combination of Python's \texttt{request} module and the BeautifulSoup HTML parser.  Mining took place between July 9 and July 12, 2024, yielding 760,460 mined models. We also used the Hugging Face API to gather the list of all datasets publicly hosted on the platform and available through the API as of July 9, 2024.  For each model in our dataset, information on base models and datasets was obtained, when available, from the \cardData attribute of the model metadata as well as from the model tags.  We make all the raw data acquired from mining available in our replication package \cite{anonymous_repo}. 

Although there are a handful of research datasets for Hugging Face models and datasets already exist \cite{jiang2024peatmoss, jiang2023ptmtorrent}, we opted to create our own dataset for this work for three reasons:
\begin{enumerate}
    \item We aimed to analyze the most current data, which includes new models and model families (\eg new models recently added to the Llama family of models) and changes in licensing frequencies (see \Cref{sec:licensing}). The presence of these new models is significant given that many of these were the first open-weight models capable of competing with the, until then, assumed more powerful and untouchable proprietary options (\eg OpenAI's Chat-GPT, Google's Gemini, and Anthropic's Claude).  The practical utility of these new models represented a paradigm shift for fine-tuning, reusing, and remixing existing open-weight models.  Their inclusion is essential to understand the present state of the ML supply chain on Hugging Face.%
    \item We aimed to collect and analyze a more comprehensive and complete dataset; as shown in \Cref{tab:dataset-comparison}, our dataset is significantly larger than most existing datasets and comprises both datasets and models.
    \item Unlike Jiang \etal \cite{jiang2024peatmoss} and Pepe \etal \cite{pepe2024hugging}, we were not focused on model-to-GitHub repository dependencies; accordingly, we aimed to design a dataset tailored for mapping model-to-model and dataset-to-model dependencies.
\end{enumerate}

\begin{table}[]
\caption{Comparison of study goals and Hugging Face (HF) datasets used in prior work} %
\label{tab:dataset-comparison}
\resizebox{\columnwidth}{!}{
\begin{tabular}{|crrc|}
\hline
\textbf{Work} & \multicolumn{1}{l}{\textbf{Models}} & \textbf{Datasets} & \textbf{Study Goal(s)} \\ \hline
\cite{yang2024navigating} & - & 24,065 & Explore the state of dataset documentation on HF \\
\cite{jiang2024naming} & 500 & - & Investigate PTM* naming practices/anomolies on HF \\
\cite{jiang2023ptmtorrent} & 15,913 & - & Create a dataset for future research \\
\cite{oreamuno2024state} &  55,280 & 6,758 & Understand the state of model documentation practices \\
\cite{Jiang2023AnES} & 63,182 & - & Understand model reuse, developer workflow, and challenges \\
\cite{pepe2024hugging} & 159,132 & - & Study extent to which model maintainers document PTMs \\
\cite{taraghi2024deep} & 239,422 & - & Understand challenges faced during PTM reuse \\
\cite{jiang2023peatmoss} / \cite{jiang2024peatmoss} & 281,638 & - & Create a dataset examining relationship between PTMs and OSS \\
\cite{castano2024lessons} / \cite{castano2024analyzing} & 380,000 & - & Provide guide for researchers/explore how models are maintained \\
\textbf{This work} & \textbf{760,460} & \textbf{175,000} & Explore structure of ML supply chain and licensing challenges \\ \hline
\end{tabular}
}%
\\
\footnotesize{*Pre-Trained Model}
\end{table}

\subsection{Data normalization and cleaning}
Once the data for models and datasets were downloaded, the data needed to be cleaned, standardized, and stored in a more usable form.  This process involved standardizing the names of declared base models and datasets 
and mapping to unique identifiers where possible. 
For example, \model{xlm-roberta-base} is a shorthand for \model{FacebookAI/xlm-roberta-base}, which has the unique internal ID \textsc{621ffdc036468d709f174364}. We relied on Hugging Face's internal mechanisms to resolve short names where possible. %
For example, attempting to load the model page for \model{xlm-roberta-base} on Hugging Face redirects to the page for \model{FacebookAI/xlm-roberta-base}. Additionally, we resolved inconsistencies in field inputs. For example, [], ``'', and \textsc{None} were all being used to denote no declared licenses. We standardized these variations to the empty list to make later computations simpler. %
We further elaborate on documentation challenges that required special attention and resolution in our discussion of \ref{rq:2}. %

\subsection{Extracting licensing information}
\label{sec:license_extraction}
We extracted licensing information from the tags present in model metadata. While licenses can also be declared in the \cardData attribute of the model metadata, we found through an analysis of our dataset that the tags were more complete in every instance than the \cardData attribute.  %
In nearly all instances where licensing information was present, it was found in both locations (99.9\%). %
However, for 0.1\% of models, the licensing information could be found exclusively in the tags, and there were no instances that relied only on the \cardData attribute.  We also looked for discrepancies in the cases where both locations were utilized.  We identified only 134  discrepancies, and in all cases, an additional license was declared in the tags that was not declared in the \cardData. As there were no irreconcilable differences (\ie the tags declared an \license{MIT} license while the \cardData attribute declared a \license{GPL} license), we concluded that the model tags are the most complete machine-readable source for licensing information. %

Although licensing information can also be found in sections of a model card that are more difficult to programmatically extract from, %
we sought to further confirm that reliance on tags to identify the declared license used by the model %
was justified. Accordingly, one author manually reviewed the model cards for the top 100 models (sorted by number of downloads) marked as having no license in the tags.  We anticipated that these 100 models were more likely to have complete and reliable documentation because of their popularity as compared with other models in our dataset, consistent with the findings of Jones \etal \cite{Jones2024WhatDW}.  From this sample of 100 models, we identified only 16 models that had declared licensing information elsewhere. Of those, one had licensing information added to its tags after our initial mining, bringing the total to 15 models. For nine of these models, the license could only be found transitively by following a link to the model's associated GitHub page.  \Cref{tab:manual_analysis} provides a more detailed breakdown of where the licensing information was located for each example.%

\begin{table}[]
\caption{License Information Location for Sample with No Tags after Manual Review}
\label{tab:manual_analysis}
\begin{tabular}{lrr}
\hline
\textbf{Location} & \multicolumn{1}{l}{\textbf{Count}} & \multicolumn{1}{l}{\textbf{Percent}} \\ \hline
Link to GitHub & 9 & 56.3\% \\
License referenced in README & 2 & 12.5\% \\
License text embedded in README & 2 & 12.5\% \\
LICENSE File & 2 & 12.5\% \\
License in tags (added after mining) & 1 & 6.3\% \\ \hline
Total & 16 & 100.0\%\\
\hline
\end{tabular}%
\end{table}

Detecting license declarations present in other sources (\eg plain text or links to external sites) %
would involve more complicated mining techniques, such as those that could traverse external links, and would require NLP-based approaches that extract licensing information from the text.  These NLP approaches would have to not only recognize license names, which may not be in a standardized form, but also pattern-match embedded license text. An LLM tool, like the one employed by Jiang \etal \cite{jiang2024peatmoss}, could potentially be used to extract licensing information, but at a high cost (between \$0.01 and \$0.03 per model---or between \$7,000 and \$22,800 for our dataset).  Relatedly, relying on a cheaper but smaller, locally hosted LLM would risk introducing hallucinated license information and so still require laborious manual human oversight. To ensure that our analysis captured the most accurate information possible and to better understand the formal documentation structure of Hugging Face, rather than employ such an approach, we focus on explicitly declared licenses.
Given the likelihood that licenses are declared in tags, the infeasibility of manual annotation for all examples, and the inherent difficulties of automatically obtaining license information from anywhere in a model card, our reliance on tags for license detection appears to be an acceptable approach. Additionally, existing supply chain analysis tools rely heavily on machine-readable metadata, so we further justify that this is the appropriate locus of analysis given that the ultimate audience for such tools is the end user. %

\subsection{Data analysis}

To address \ref{rq:2}, we quantitatively and qualitatively analyzed and discussed the different types of documentation issues that we observed in our attempt to understand the ML supply chain on Hugging Face.

Once the dataset had been cleaned, to address \ref{rq:1}, we used the \texttt{networkx} Python library \cite{networkx} to create a directed %
graph representing the observable  %
\supplychain.  Due to the inherent issues and incompleteness of model documentation, this graph, by its nature, only represents a subset of the total ML supply chain on Hugging Face. Each node in the graph represents a distinct model, and an edge represents a dependency relationship. For example, if model $A$ listed model $B$ as a base model, there would be an edge from $B$ to $A$. We would describe model $B$ as the parent of model $A$ and model $A$ as the child of model $B$. That is, model $A$ is derived from model $B$. %

For the sake of analysis, we flatten the individual model supply chains by extracting all longest paths (\ie lineage chains) %
from the graph that go from source nodes to sinks.  Each of these paths goes %
from a root base model to a model at the end of the ML supply chain (\ie a leaf node with no dependents or outgoing connections--- no other models are derived from this model).

Using the paths resulting from \ref{rq:1}, we addressed \ref{rq:3} by analyzing the state of licensing in the ML supply chain, developing scripts and heuristics to look for parent/child license differences %
and potential license compliance problems related to model/dataset dependencies.  %
We define a \textit{parent/child license difference} to refer to the difference between the license(s) of a parent and a child model/dataset. Assume, for example, that model $B$ is the child of model $A$. If model $B$ was licensed under \license{MIT} and model $A$ under \license{Apache-2.0}, we would say there is a parent/child license difference between $A$ and $B$. We avoid stronger terms such as ``inconsistencies'' and ``incompatibilities,'' since some of these differences may in fact be compatible with licensing terms. \add{This is consistent with prior work \cite{duan2024modelgo} that found that the application of licensing terms, particularly for existing licenses, in the novel ML context can introduce uncertainties and ambiguities in terms of legal interpretation.} %
Two authors reviewed all distinct licenses in our dataset and grouped them into six categories, which can provide the basis for future research into license compatibility and licensing trends in the ML supply chain. 

All scripts used for mining, data clean-up, and analysis can be found in our replication package~\cite{anonymous_repo} and can be used to foster further research in this area.

\section{RQ$_0$: Documentation challenges encountered on Hugging Face}
\label{sec:rq0}
In this section, we outline and discuss documentation issues that were observed on the Hugging Face platform while mapping connections in the observed ML supply chain.  Specifically, this preliminary RQ outlines the observed shortcomings and challenges to 1) elaborate potential pitfalls future researchers may face in mining Hugging Face and 2) to more thoroughly report documentation issues extant on Hugging Face that have been left unexplored by prior work. %

\begin{figure}[t]
\centering
\includegraphics[width=\linewidth]{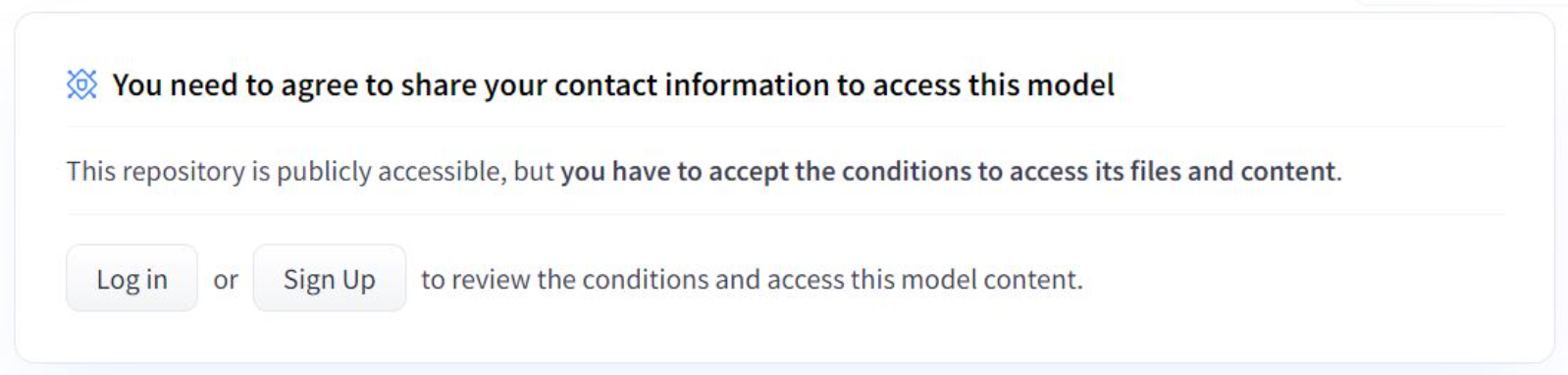}
\caption{Example of model gated behind Terms of Service requirement}
\label{fig:gated}
\end{figure}

\subsection{Inability to Access Metadata}
\label{sec:no_api_access}
Not all model metadata was accessible through the Hugging Face API.  There were 7,258 model cards (0.95\%) %
that had to be scraped using a combination of Python's \texttt{requests} \cite{requests} and BeautifulSoup \cite{bs4} libraries.  The most frequent challenge, found in 99.4\% of these cases, resulted from gating based on the acceptance of terms and conditions (see \Cref{fig:gated}).  This was particularly the case for models owned by or relying on models created by larger entities such as Google or Meta.  Twenty-two models and their metadata were gated behind age restrictions imposed by the model owner\footnote[1]{Throughout the paper, by "model owner", we refer to the entity that uploaded the model to Hugging Face.  We make no claims as to whether that entity had the legal right to do so or if they are the original author/creator of the model.}, %
requiring a Hugging Face account to access. Accessing the full repository, and thus metadata, required logging in with a Hugging Face account and then, in most cases, providing an email address.  These hurdles to accessing the metadata for a model could potentially complicate the creation of fully automated analyzers and ML/AIBOM (Machine Learning/Artificial Intelligence Bill of Materials)~\cite{stalnaker2024boms, xia2023empirical, barclay2019towards, barclay2022providing} %
generators for an ML supply chain\add{, which we believe should not rely on scraping potentially changing HTML content or on Selenium-based web-drivers for form completion and navigation.}

\subsection{Incomplete Metadata}
\label{sec:incomplete_metadata}
We saw, %
consistent with prior work \add{\cite{oreamuno2024state,Jiang2023AnES,jiang2024peatmoss, jiang2023ptmtorrent}}, that the documentation provided by model owners on Hugging Face is often incomplete. %
This is a trend not just for smaller, infrequently used projects but also for large projects with tens of thousands of downloads. For example, OpenAI's \model{clip-vit-large-patch14-336} model, with 5,827,027 downloads, has a model card that is almost entirely incomplete, save for basic hyper-parameter and framework version information  \cite{incomplete_docs_example}.  %
Only 37.8\% of models and 27.6\% of datasets declared any kind of licensing information in a standardized, machine-readable way.  We discuss the prevalence of declared components further in \Cref{sec:rq1}. %

Some models may provide information on training data and architecture by linking to scientific papers that include these details, but such a practice introduces an additional hurdle in easily and programmatically obtaining that data, contributing to the ML reproducibility crisis~\cite{jiang2024challenges}.  Ultimately, the reason why many models do not even mention the datasets used would require further investigation, including interviewing/surveying developers, which is out of the scope of this work.   %

During the manual analysis of 100 model cards described in \Cref{sec:license_extraction}, we observed two examples %
where the Hugging Face team wrote documentation on behalf of model owners.  The first instance was an image-to-text model  uploaded by Microsoft with the disclaimer: ``The team releasing TrOCR did not write a model card for this model so this  model card has been written by the Hugging Face team'' \cite{hugging_face_authored_1}. The second was a larger version of
the same model, also owned by Microsoft \cite{hugging_face_authored_2}.  In both cases, the models were described by a research paper and released on GitHub.  Understanding the criteria that the Hugging Face team uses to determine when and  how to intervene with model cards requires further investigation, but we do note that according to a blog written by 
members of the Hugging Face team, they created/updated model cards in some instances to inform design decisions surrounding a new 
model card template~\cite{hugging_face_model_card_blog}. %

\subsection{The \unknown license}
The Hugging Face platform's documentation provides a list of recognized licenses each with a unique, standardized short-hand identifier that can be utilized by users when creating their model/data cards~\cite{hugging_face_licenses}. %
This list includes common Open Source Software (OSS), Creative Commons (CC), and ML-specific licenses as well as an \other catch-all category that encompasses less common, custom, or modified licenses not included in Hugging Face's predefined license list. %
An \unknown licensing option is also provided, but there is no guidance provided as to when this option should be selected. Future work will need to explore when and why model owners select this option, particularly given that acknowledging that the license is \unknown suggests, at best, a lack of due diligence or understanding on the part of the owner or, at worst, that the model/dataset raises copyright infringement issues.  We observed 4,419 (1.5\%) models and 2,194 (4.5\%) datasets that used this \unknown license tag, which inherently raises compliance challenges for dependent models.
\looseness=-1

\subsection{Naming problems}
\label{sec:naming-problems} 
Hugging Face users primarily use human-readable names, not unique identifiers, when supplying the metadata information for their base models/datasets.  Unique identifiers do exist for all models and datasets on Hugging Face, but we did not observe any instances where this information was being used by model/dataset owners when referring to models or datasets they depended on. Instead, these references typically, but not always, followed the ``owner/model'' naming convention. %
For example, a model owner would refer to a base model as ``FacebookAI/xlm-roberta-base'' and not by its unique Hugging Face ID: 621ffdc036468d709f174364. This is problematic because these model references are not automatically updated when name changes occur. A change to the model, dataset, or owner name results in a new human-readable identifier that no longer matches the previous references to the model/dataset.  As a result, documentation can easily become outdated, unhelpful, and potentially confusing.

We observed 596 models that had their human-readable identifiers changed. %
Instances such as \model{222gate/Blurdus-7b-v0.1} $\rightarrow$ \model{gate369/Blurdus-7b-v0.1} involve a change to the name of the model owner.  Other instances, such as \model{aaditya/openbiollm-llama3-70b} $\rightarrow$ \model{aaditya/Llama3-OpenBioLLM-70B}, involve a change to the model name itself. In both cases, manual effort is required to track down and map the new names to the original references, which can make dependency management tasks increasingly difficult.  Often the only way to determine the correct mapping is by relying on Hugging Face to redirect the page associated with a former name to the one associated with the most current name.  This suggests that Hugging Face maintains some internal mapping, but it is unclear how long this mapping persists or what might happen in the event of a name collision.

\begin{table}[]
\caption{Top 10 Most Common Model Names in the Ecosystem}
\label{tab:common_names}
\begin{tabular}{lr}
\hline
\textbf{Model Name} & \textbf{Count} \\ \hline
ppo-LunarLander-v2 & 5,660 \\
q-FrozenLake-v1-4x4-noSlippery & 4,200 \\
test & 3,900 \\
ppo-Huggy & 2,684 \\
dqn-SpaceInvadersNoFrameskip-v4 & 2,405 \\
distilbert-base-uncased-finetuned-emotion & 2,348 \\
my\_awesome\_model & 1,727 \\
q-Taxi-v3 & 1,535 \\
ppo-SnowballTarget & 1,513 \\
bert-finetuned-ner & 1,297\\
\hline
\end{tabular}%
\end{table}

Model names, without the additional owner information, are not guaranteed to be unique.  Similar to how repository forks have the same name (but different owners) on GitHub, fine-tuned or forked models may also have the same name yet different owners on Hugging Face.  For this reason, developers often specify models using the ``owner/model'' naming convention described previously. If this happens consistently, ambiguities can be avoided. However, we observed 16,477 cases where developers referred to base models only by the model name, excluding owner information. For example, a developer might use ``roberta-base'' to refer to ``FacebookAI/roberta-base.''  While it could be reasonable to assume that this shorthand refers to the popular model, this may not be the case: in our dataset, there were 39 other models also named ``roberta-base,'' all with different owners.  \Cref{tab:common_names} shows the top 10 most common model names in the Hugging Face ecosystem and their frequencies.  If one of these models were referred to by name only, it would be nearly impossible to determine which model was being referenced. For example, more than 5,600 models share the model name ``ppo-LunarLander-v2.''  While many of these are likely toy models or ``Hello World'' equivalents, others, such as \model{distilbert-base-uncased-finetuned-emotion}, might suggest the need for widespread adoption of unique model identifiers. %

\subsection{Missing or nonsensical references} 
\label{sec:missing_refs}
There were 34,159 instances where we were unable to map a declared dataset (\eg a dataset that a model reported relying on for training) to a dataset found publicly on Hugging Face. %
These consisted of 4,755 unique declarations.  It is impossible to know with certainty, but such declarations likely consist of datasets that have been removed or made private, are from external sources, are actually dataset descriptors, contain typos, or, in some cases, are Hugging Face usernames. 

We were also unable to map 2,501 base model declarations to known IDs, representing a possible 1,371 unique missing models in total. %
As with the datasets, these models may have been removed, made private, renamed in a way difficult to trace, or may reference a model outside the ecosystem.  It is also possible that developer typos result in the inability to identify the models. 

In other cases, some of the references were nonsensical. In our analysis of the graph of all collected components in the ML supply chain (described further in \Cref{sec:rq1}), we detected 684 cycles in total, including 675 (98.7\%) trivial cycles in which a model declares itself as a base model, despite the fact that a model should not be able to be its own ancestor. All of this further motivates the need to use unique and standardized identifiers when referring to models and datasets, as well as a need for validating reference information.

\subsection{Models as datasets} \label{sec:models_as_data} We observed 1,416 instances in which a model was listed as a dataset in the \textsc{datasets} field. In other words, a model reference was included in the metadata field specific to training datasets.  Without more information, it is impossible to know whether this indicates a mistake on the part of the model owner or that the output of the declared model was somehow used for training. 

\subsection{Shortcomings of Hugging Face} 
\label{sec:no_relationship}
While base models are declared by some model owners, in many cases, the precise relationship between a derived model and its base model is left ambiguous, at least in the available metadata. Hugging Face does provide a separate field for architecture relationships, but there is no standardized way to specify situations involving fine-tuning, quantization, or using outputs for training.  This ambiguity can have significant consequences since, particularly in the licensing context, the nature of the relationship is important in determining how licensing terms should be applied. For example, it is permissible to use a model licensed under the \license{llama3} license for fine-tuning but impermissible to use its outputs to train a competing model~\cite{llama3_license}.  Additional fields specifying the nature of relationships between models would be useful, but their inclusion would also introduce additional administrative burden for model owners.

While uncommon and not a direct shortcoming of Hugging Face, we observed examples where model owners would declare the same dataset or base model multiple times (\ie the same model/dataset had multiple entries in the respective metadata field). Specifically, 178 models declared a dataset at least twice, and 310 models declared a base model at least twice.  One model, with otherwise robust documentation, declared the same base model 64 times. Again, while there are few instances of this duplication, they still serve as something of a canary in the coal mine, indicating a lack of validation on behalf of Hugging Face.

\subsection{Summary}

In summary, we identified numerous challenges while gathering data from Hugging Face, ranging from incomplete model information to the inability to access existing data. We highlight the prevalence of these issues to alert others seeking to mine Hugging Face and to identify specific areas for improvement, both from those documenting their models on Hugging Face and for any future efforts to improve Hugging Face's system.

\section{RQ$_1$: Analysis of the ML supply chain}
\label{sec:rq1}

\begin{table}[t!]
\caption{Top 10 Models by Likes}%
\label{tab:top_models_likes}
\begin{tabular}{llrl}
\hline
\textbf{Model} & \textbf{Pipeline} & \textbf{Likes} & \textbf{License} \\ \hline
runwayml/stable-diffusion-v1-5 & text-to-image & 10,987 & creativeml-openrail-m \\
CompVis/stable-diffusion-v1-4 & text-to-image & 6,358 & creativeml-openrail-m \\
stabilityai/stable-diffusion-xl-base-1.0 & text-to-image & 5,502 & openrail++ \\
meta-llama/Meta-Llama-3-8B & text-generation & 5,200 & llama3 \\
bigscience/bloom & text-generation & 4,647 & bigscience-bloom-rail-1.0 \\
meta-llama/Llama-2-7b & text-generation & 3,980 & llama2 \\
mistralai/Mixtral-8x7B-Instruct-v0.1 & text-generation & 3,960 & apache-2.0 \\
stabilityai/stable-diffusion-2-1 & text-to-image & 3,778 & openrail++ \\
WarriorMama777/OrangeMixs & text-to-image & 3,714 & creativeml-openrail-m \\
meta-llama/Llama-2-7b-chat-hf & text-generation & 3,670 & llama2 \\ \hline
\end{tabular}%

\bigskip

\caption{Top 10 Models by Downloads}
\label{tab:top_models_downloads}
\resizebox{\columnwidth}{!}{
\begin{tabular}{llrl}
\hline
\textbf{Model} & \textbf{Pipeline} & \textbf{Downloads} & \textbf{License} \\ \hline
MIT/ast-finetuned-audioset-10-10-0.4593 & audio-classification & 677,161,532 & bsd-3-clause \\
sentence-transformers/all-MiniLM-L12-v2 & sentence-similarity & 83,805,061 & apache-2.0 \\
facebook/fasttext-language-identification & text-classification & 53,517,573 & cc-by-nc-4.0 \\
google-bert/bert-base-uncased & fill-mask & 52,724,375 & apache-2.0 \\
sentence-transformers/all-MiniLM-L6-v2 & sentence-similarity & 41,281,684 & apache-2.0 \\
openai/clip-vit-large-patch14 & zero-shot-image-classification & 40,114,370 & - \\
distilbert/distilbert-base-uncased & fill-mask & 24,195,717 & apache-2.0 \\
openai/whisper-small & automatic-speech-recognition & 23,317,804 & - \\
jonatasgrosman/wav2vec2-large-xlsr-53-english & automatic-speech-recognition & 18,212,315 & apache-2.0 \\
openai/clip-vit-base-patch16 & zero-shot-image-classification & 16,999,162 & - \\ \hline
\end{tabular}%
}
\end{table}

Despite the challenges outlined in \ref{rq:2}, we were still able to construct an ecosystem-level supply chain graph mapping the relationships between models and datasets. While we acknowledge that our supply chain graph represents only a subset of the total components %
in the full Hugging Face ecosystem, as noted previously, %
we believe that it is appropriate to rely on it to make observations about recurring trends and patterns. %

To analyze the structure of the ML supply chain on Hugging Face, we began by addressing the issues identified in \ref{rq:2} using different strategies depending on the issue. %
Where possible, we employed web-scraping techniques to pull metadata for models that were inaccessible through the Hugging Face API (\Cref{sec:no_api_access}).  Models where even this analysis was impossible were excluded from the analysis.  We standardized model/dataset names, preferring the unique Hugging Face ID (\Cref{sec:naming-problems}).  Where naming ambiguities were present (multiple models with the same name, but different owners), we assumed the most popular model (\Cref{sec:naming-problems}).  References that could not be mapped to existing models/datasets on Hugging Face were also excluded from analysis (\Cref{sec:missing_refs}).  Models that were listed as datasets were excluded from the dataset analysis (\Cref{sec:models_as_data}). %

\subsection{Most popular models and datasets}
\label{ssec:popular_models_datasets}
\begin{table}[t]
\caption{Top 10 Datasets by Likes} %
\label{tab:top_datasets_likes}
\resizebox{\columnwidth}{!}{
\begin{tabular}{lllrl}
\hline
\textbf{Dataset} & \textbf{Declared task} & \textbf{Size} & \textbf{Likes} & \textbf{License} \\ \hline
fka/awesome-chatgpt-prompts & Question Answering & <1K & 5,011 & cc0-1.0 \\
HuggingFaceFW/fineweb & Text Generation & 10B-100B & 1,539 & odc-by \\
Open-Orca/OpenOrca & Text Classification & 1M-10M & 1,269 & mit \\
OpenAssistant/oasst1 & - & 10K-100K & 1,232 & apache-2.0 \\
gsdf/EasyNegative & - & <1K & 1,128 & other \\
Anthropic/hh-rlhf & - & 100K-1M & 1,092 & mit \\
togethercomputer/RedPajama-Data-1T & Text Generation & 1M-10M & 1,025 & - \\
Nerfgun3/bad\_prompt & - & <1K & 921 & creativeml-openrail-m \\
tiiuae/falcon-refinedweb & Text Generation & 100M-1B & 765 & odc-by \\
allenai/dolma & Text Generation & >1T & 752 & odc-by \\ \hline
\end{tabular}%
}

\bigskip

\caption{Top 10 Datasets by Downloads}
\label{tab:top_datasets_downloads}
\resizebox{\columnwidth}{!}{
\begin{tabular}{lllrl}
\hline
\textbf{Dataset} & \textbf{Declared task} & \textbf{Size} & \textbf{Downloads} & \textbf{License} \\ \hline
hails/mmlu\_no\_train & Question Answering & -* & 16,146,384 & mit \\
lighteval/mmlu & Question Answering & 1M-10M & 5,882,652 & mit \\
argilla/databricks-dolly-15k-curated-en & - & 10K-100K & 4,671,897 & - \\
ceval/ceval-exam & Text Classification & 10K-100K & 2,111,340 & - \\
lavita/medical-qa-shared-task-v1-toy & - & <1K & 1,792,638 & - \\
lukaemon/bbh & - & 1K-10K & 911,736 & - \\
haonan-li/cmmlu & Multiple Choice/Question Answering & 10K-100K & 633,608 & cc-by-nc-4.0 \\
chansung/requested-arxiv-ids-3 & - & <1K & 587,249 & - \\
cais/mmlu & Question Answering & 100K-1M & 555,115 & mit \\
allenai/ai2\_arc & Question Answering & 1K-10K & 543,678 & cc-by-sa-4.0 \\ \hline
\end{tabular}%
}
{\footnotesize * No size is provided since the dataset requires Python code execution}
\end{table}

We begin with an overview of the landscape by describing the most popular models and datasets. We measure popularity in terms of engagement via the metrics provided by Hugging Face: likes and recent downloads. These metrics also serve our goal of capturing the current state of the ecosystem by highlighting what models and datasets are currently being engaged with. %
\Cref{tab:top_models_likes,tab:top_models_downloads} show the ten most engaged with models in our dataset by these metrics. Organizations such as OpenAI and Meta occupy several top positions. The models span a variety of use cases, including image and text generation.

Similarly, \Cref{tab:top_datasets_likes,tab:top_datasets_downloads} display the most engaged with datasets. The most common \emph{Task} field is ``Question Answering'' for the most downloaded datasets  and ``Text Generation'' for the most liked datasets. However, a variety of \emph{Data Modalities} and \emph{Tasks} are represented across both groups. For example, \emph{Data Modalities} included both text and images, while \emph{Tasks} included other specifics such as ``Token Classification.'' In fact, the most liked dataset by far, \model{fka/awesome-chatgpt-prompt},
is a repository of curated ChatGPT prompts rather than a training dataset for models, which suggests that it may be more prudent to examine download count to determine the most popular datasets. This is further supported by the download count's correlation with the number of client projects on GitHub \cite{pepe2024hugging}.%

\subsection{Most depended upon datasets and base models}
\label{sec:most-dependend-upon}

Our ability to evaluate the ML supply chain relies on the dependency information provided by model owners.  On Hugging Face, this information is found in the model metadata, where model owners declare base models and datasets. Although many models on Hugging Face were not created independently, we note that only 117,245 out of 760,460  models (15.4\%) declare any base model and, of these, most (111,568 | 95.2\%) declare only one base model. Overall, most models do not declare any base models: with respect to the number of base models declared, the average, as well as the minimum, first quartile, median, and third quartile, are all 0. The maximum is 53 unique base models. %

\begin{table*}[b]
\caption{Top 10 Declared Base Models}%
\resizebox{\columnwidth}{!}{
\label{tab:top_base_models}
\begin{tabular}{lrrrll}
\hline 
\textbf{Base Model}       & \textbf{Times Used} & \textbf{Likes}  & \textbf{Downloads} & \textbf{Pipeline} & \textbf{License}  \\ \hline
distilbert/distilbert-base-uncased       & 5,240      & 463    & 24,195,717 & fill-mask & apache-2.0            \\
stabilityai/stable-diffusion-xl-base-1.0 & 4,776      & 5,502  & 3,558,713  & text-to-image & openrail++            \\
runwayml/stable-diffusion-v1-5           & 2,438      & 10,987 & 3,790,115  & text-to-image & creativeml-openrail-m \\
openai-community/gpt2                    & 2,405      & 2,076  & 6,888,015  & text-generation & mit                   \\
unsloth/llama-3-8b-bnb-4bit              & 2,363      & 141    & 486,478    & text-generation & llama2                \\
FacebookAI/xlm-roberta-base              & 2,169      & 526    & 6,196,387  & fill-mask & mit                   \\
mistralai/Mistral-7B-v0.1                & 2,099      & 3,290  & 706,043    & text-generation & apache-2.0            \\
google-bert/bert-base-uncased            & 1,576      & 1,665  & 52,724,375 & fill-mask & apache-2.0            \\
google-bert/bert-base-cased              & 1,472      & 235    & 5,191,167  & fill-mask & apache-2.0            \\
openai/whisper-small                     & 1,469      & 180    & 23,317,804 & speech-recognition & apache-2.0            \\ \hline
\end{tabular}
}
\end{table*}

\Cref{tab:top_base_models} lists the ten models that were most frequently declared as a base model by the models in our dataset, \ie the models that were most frequently cited as dependencies by other models. They exhibit a varying number of likes and downloads and primarily come from large providers such as Meta, Google, and OpenAI. Notably, none of these models declare that they are derived from any base models. We observe some overlap with the ten most engaged with models by downloads%
, suggesting, not surprisingly, that popular models are also frequently used as base models. %

In addition to base models, we investigate the datasets used for model training. We observe that 75,516 (9.9\%) models declare at least one dataset. Of these, 64,343 (85.2\%) declare a single dataset, 5,066 (6.7\%) declare two datasets, and 1,637 (2.2\%) declare three datasets. The number of datasets also exhibits long-tail behavior, with one model (\model{tasksource/deberta-base-long-nli}) declaring 287 distinct datasets. %
However, the vast majority of models declare no dataset, with 0 declared models as the average, minimum, and first three quartiles, while the maximum is 287 datasets. %
We observe that this low percentage of declared datasets is particularly notable since, by their nature, nearly all ML models on Hugging Face should have been trained on at least one dataset, even if that dataset is not publicly available. %

\begin{table*}[]
\caption{Top 10 Declared Datasets}
\label{tab:top_declared_datasets}
\resizebox{\columnwidth}{!}{%
\begin{tabular}{lrrrl}
\hline
\textbf{Dataset} & \textbf{Times Used} & \textbf{Likes} & \textbf{Downloads} & \textbf{License} \\ \hline
nyu-mll/glue & 2,293 & 336 & 122,473 & other \\
rajpurkar/squad & 1,827 & 228 & 11,379 & cc-by-sa-4.0 \\
dair-ai/emotion & 1,766 & 257 & 23,236 & other \\
ilsvrc/imagenet-1k & 1,470 & 345 & 17195 & other \\
mozilla-foundation/common\_voice\_7\_0 & 1,330 & 163 & 67,565 & cc0-1.0 \\
stanfordnlp/imdb & 1,329 & 208 & 20,078 & other \\
legacy-datasets/wikipedia & 1,219 & 528 & 1,657 & cc-by-sa-3.0, gfdl \\
google/xtreme & 1,154 & 82 & 982 & \begin{tabular}[c]{@{}c@{}}apache-2.0, cc-by-4.0, cc-by-2.0,\\ cc-by-sa-4.0, other, cc-by-nc-4.0\end{tabular} \\
mozilla-foundation/common\_voice\_11\_0 & 1,130 & 163 & 67,565 & cc0-1.0 \\
eriktks/conll2003 & 880 & 115 & 6,171 & other \\ \hline
\end{tabular}%
}
\end{table*}

\Cref{tab:top_declared_datasets} describes the top ten declared datasets. We observe no overlap with the ten most liked or downloaded datasets.  This may be a consequence of the low declaration rate for datasets, a function of a dataset's age, or the fact that popular ``datasets'' uploaded to Hugging Face may not be intended for model training, as observed above.

License declarations vary among the datasets. Half of the ten most declared datasets are made available under a license declared as ``Other.'' One example from Google appears under six distinct licenses, including the \license{apache-2.0} license, several Creative Commons licenses, and an \other license. While these are all listed in the dataset's metadata, the dataset card's field for licensing information simply says, ``More Information Needed,'' leaving it unclear as to how these licenses jointly apply to the dataset. This suggests that even very popular components can contain ambiguities that make the license compliance process more difficult. %

\subsection{Structure of the supply chain}

As explained in \Cref{ssec:data_collection}, our mined dataset consists of over 760,000 models. Our final model supply
chain graph, which features the reported dependency relationships between models, features
761,826 nodes (models) and 126,995 directed edges. %
We remove any edges related to trivial cycles discussed in Section 4.3.3 where a model reports itself as its own base. This leaves us with
126,320 directed edges. %
The number of models in the graph exceeds the number in the dataset because some models were declared as dependencies but were not observed on Hugging Face at the time of mining. These models might have been removed, made private, exist on another platform, or had a name change that we were unable to trace.

The constructed supply chain graph can provide valuable information to model owners and researchers. For example, we can determine which models in the Hugging Face ecosystem
are the most relied upon, and thus for which models complete, accurate documentation, compliance checks, and security audits may be more critical. %

\subsubsection{Analysis of Removed Cycles}

As noted above, we removed from the analysis 675 trivial cycles. Of these, 168 featured nodes that were connected to other models in the supply chain graph.  There were 154 instances (92\%) where a model reportedly depended on itself and was also reported as a base for a another model.  There were 14 instances (8\%) in which a model reportedly depended on itself and at least one other model. In most cases, these cycles likely represent typos introduced by the model owner, but they could
potentially also be artifacts indicating iterative fine-tuning or some other additional training.

\subsubsection{Analysis of graph degree and density}

There were 635,966 models (83.5\%) isolated from the rest of the graph with no incoming or outgoing edges, representing the vast majority. In other words, they have a degree of zero. The next largest portion of nodes, a cohort of 111,840 (14.7\%), have degree one. (``Degree'' here refers to the total count of incoming \textit{and} outgoing edges.) %
\Cref{tab:graph_degree} shows the breakdown for statistics for up to degree five. There was one model, \model{distilbert/distilbert-base-uncased}, that had an astounding degree of 5,240. For this model, all of these represented outgoing edges, meaning that 5,240 models recorded depending on \model{distilbert/distilbert-base-uncased}. Taking all this together, the graph’s density is very low ($2.18 x 10^{-7}$), reinforcing that most models do not record any model dependency relationships.  We rely on \library{networkx}'s \library{density} function to calculate and return the graph density, using \Cref{eq:density}.

\begin{equation}
\label{eq:density}
D = \frac{|E|}{|V|(|V| - 1)}
\noindent
\textit{, where $|V|$ is the number of nodes and $|E|$ is the number of edges.}
\vspace{1em} %
\end{equation}

\begin{table}[]
\caption{Nodes of Degree \textit{N} in Model Graph}
\label{tab:graph_degree}
\begin{tabular}{lll}
\hline
\textbf{Degree*} & \textbf{Node Count} & \textbf{Percentage} \\ \hline
0                                     & 635,966                             & 83.5\%                                   \\
1                                     & 111,840                             & 14.7\%                                   \\
2                                     & 6,643                               & 0.9\%                                    \\
3                                     & 2,550                               & 0.3\%                                    \\
4                                     & 1,392                               & 0.2\%                                    \\ 
5                                     & 764                                 & 0.1\%                 \\ \hline                  
\end{tabular}
{\\ \footnotesize * Degree refers to the total count of incoming and outgoing edges}
\end{table}

There are 13,985 model nodes %
that have at least one outgoing edge, meaning they are depended upon by at least one other model. Almost half of these (7,032) have at least two outgoing edges. 620 model nodes have an outgoing degree of twenty or more, and 150 have an outgoing degree of one hundred or more. There are 116,576 model nodes that have at least one incoming edge, meaning they record depending on at least one base model. Only 5,673 of these records are based on two or more base models. All of this demonstrates, unsurprisingly, that models are more likely to rely on other models than they are to be relied on themselves. We also note that this shows that the machine learning ecosystem is upheld by more than just a handful of well-known and readily recognizable models, such as the 150 models with one hundred or more direct dependents.

\subsubsection{Connected components (subgraphs)}

There are 643,653 connected components (\ie disjoint subgraphs) within our supply chain graph. %
Nearly all of these, 635,966 (98.8\%), are trivial subgraphs including just a single node. An additional 3,550 subgraphs are made up of only two nodes. \Cref{tab:subgraph_order_freq} breaks down the most common order frequencies for the subgraphs. From this, it is clear that the connected components constructed based on available documentation tended to be small, but there were 107 subgraphs that contained one hundred or more nodes.%

\begin{table}[]
\caption{Subgraph Order Frequencies}
\label{tab:subgraph_order_freq}
\begin{tabular}{ll}
\hline
\textbf{Order} & \textbf{Count} \\ \hline
1                                            & 635,966        \\
2                                            & 3,550          \\
3                                            & 1,349          \\
4                                            & 803            \\
5                                            & 519            \\ \hline
\end{tabular}
\end{table}

\subsubsection{An analysis of the largest subgraph}

The largest subgraph featured 28,879 nodes and 35,633 edges, giving it a density of $8.54x10^{-5}$. The majority of nodes in this graph, 22,099 (76.5\%), have a degree of one. %
This is followed by 3364 (11.6\%) nodes with degree two and 1235 (4.3\%) nodes with degree three. There were multiple hubs within the supply chain that served to connect the various nodes. For example, there were 81 nodes that had a degree greater than fifty. These represent popular models that have many models fine-tuned or based on them. See \Cref{tab:top_base_models} for details on the top ten such models. %
Ultimately, we see that the sparse nature %
that is representative of the entire Hugging Face supply chain graph also applies to its largest connected component.%

\subsection{Model lineages}

\subsubsection{Lengths of model supply chains}

We define a lineage chain as a path from some root base model (\ie one with no model dependencies) to a final sink node model (\ie one without any model dependents).  We found and examine 53,151 lineage chains for models that declare at least one base model, including cases where multiple chains lead to the same model. Of these, the average chain length to reach that model is 6.2 models. The most frequent chain length is three models (12,480 chains); %
the minimum is 2 models, the first quartile is 3, the median is 4, the third quartile is 9, and the maximum is 40. %

\subsubsection{Licenses present in model supply chains}

There are typically relatively few parent/child license differences %
between models in these chains. We remind the reader than a parent/child license difference does not necessitate an incompatibility, but instead demonstrates that some licensing decisions (or lack thereof in the case of neglecting to license a derivative work) were made along the chain.  There was an average of 1.9 distinct licenses in a given chain, including unknown or undeclared licenses. %
The minimum and first quartile are both just 1 license, both the median and third quartile are 2 licenses, and the maximum is 6 licenses. %
A chain most commonly contains 2 licenses (24,069 chains), and the highest number of licenses observed was 6, occurring in 17 different chains. A relatively small number of parent/child license differences simplifies the task of managing license compliance. However, these differences within a lineage necessitate examination of license compatibility, as well as a possible need to find mitigations for items in model chains when required by the license of a base model.%

\subsubsection{Model ownership}
\label{sssection:model_ownership}

We also observe that when choosing models to build off of, developers using Hugging Face often use models from the community rather than using their own previous work as a base.
The owners of nodes at the end of a chain appeared within that node's chain only 1.4 times on average, showing that base models are frequently sourced from the community and supporting the idea that a common approach to training a model involves building on previous work. Alternatively, it could be that model trainers do not provide base model information when building on their own work at least some of the time. Future work involving discussions with developers could shed light on the reasons behind this phenomenon. %

In 82.9\% of chains (43,507), the owner of the final model in the chain owned no other model in that chain. In the maximum case, the chain contained 39 models from the final model's owner. %
The minimum, first quartile, median, and third quartile are all a single occurrence of the final model's owner in a given chain: only the final model itself.

Based on our observations, %
a given owner publishes relatively few models to Hugging Face, owning, on average, 4 models in our dataset, though the data is skewed heavily toward relatively few highly-prolific owners. Of 190,136 distinct model owners in our dataset, 61.5\% owned just one public model (117,016), while the most prolific account owned 4,610 models. The first quartile and median are both just 1 owned model, and the third quartile is 2 models. Notably, the large players mentioned above, including Meta and OpenAI, are not present in the top 10 most prolific model owners: those spots instead go to smaller stakeholders in the AI market. The top ten model and dataset owners on Hugging Face are shown in \Cref{tab:top-owners}. There is only one overlap between the two lists. 

\begin{table}[]
\caption{Top 10 model and dataset owners on Hugging Face}
\label{tab:top-owners}
\begin{tabular}{lrllr}
\multicolumn{2}{c}{\textbf{Models}}   & \multicolumn{1}{c}{\textbf{}} & \multicolumn{2}{c}{\textbf{Datasets}}                          \\ \cline{1-2} \cline{4-5} 
\textbf{Owner} & \textbf{Model Count} &                               & \textbf{Owner}                        & \textbf{Dataset Count} \\ \cline{1-2} \cline{4-5} 
mradermacher   & 4,610                &                               & open-llm-leaderboard-old              & 7,059                  \\
thebloke       & 3,863                &                               & cyberharem                            & 3,384                  \\
huggingtweets  & 3,830                &                               & autoevaluate                          & 3,232                  \\
tkcho          & 3,810                &                               & recag                                 & 2,345                  \\
jeevesh8       & 3,723                &                               & liuyanchen1015                        & 1,558                  \\
prunaai        & 3,411                &                               & adapterocean                          & 810                    \\
lonestriker    & 3,241                &                               & tyzhu                                 & 807                    \\
sail-rvc       & 3,182                &                               & joey234                               & 705                    \\
larryaidraw    & 3,046                &                               & lots-of-loras                         & 500                    \\
cyberharem     & 2,978                &                               & results-sd-v1-5-sd-v2-1-if-v1-0-karlo & 462                    \\ \cline{1-2} \cline{4-5} 
\end{tabular}
\end{table}

\subsection{Summary}

Our findings show that Hugging Face components rarely declare datasets, generally have short chains of related components, and when dependency relationships between components exist, licensing information usually stays the same from parent to child. This perhaps simplifies the task of managing dependencies and licensing, but largely by virtue of the fact that relevant supply chain information appears to be missing in many such cases.

\section{RQ$_2$: licensing of models and datasets on Hugging Face}
\label{sec:licensing}
Only 37.8\% of models and 27.6\% of datasets declared licensing information in a machine-readable way (see \Cref{sec:license_extraction}). In this section, we investigate this licensing information to better understand how datasets and models are licensed in the \supplychain. %
We begin by organizing the licenses that we observed into six categories, intended to be indicative of the origin or purpose of the licenses (whether the licenses were designed for open source software, for ML components specifically, or for other purposes) and are not intended to convey any legal characteristics, particularly given the uncertainty involved in applying these licenses in the ML context. These categories are not aimed at resolving compliance issues but instead demonstrate the complex and diverse landscape of licenses in the \supplychain. %
We define the following categories: %
\begin{itemize}
    \item OSS: open source software licenses such as \license{MIT}, \license{Apache}, and \license{GPL}
    \item CC: Creative Commons licenses
    \item ML: ML-specific licenses such as \license{open-rail} and \license{llama}
    \item Data: licenses that are intended to specifically protect data such as \license{odbl} and \license{c-uda}
    \item Other: the \other license category on Hugging Face %
    \item Unknown: the \unknown license category on Hugging Face
\end{itemize}

\subsection{Most common licenses}
\label{sec:common_licenses}

\begin{table}[]
\caption{Top 10 licenses for models.}
\label{tab:top_licenses_models}
\begin{tabular}{lcrr}
\multicolumn{4}{c}{\textbf{Models}} \\ \hline
\textbf{License} & \textbf{Type} & \textbf{Count} & \textbf{\begin{tabular}[c]{@{}r@{}}Percent \\ (of licensed)\end{tabular}} \\ \hline
apache-2.0 & OSS & 119,449 & 41.6\% \\
mit & OSS & 51,184 & 17.8\% \\
openrail & ML & 30,095 & 10.5\% \\
creativeml-openrail-m & ML & 20,396 & 7.1\% \\
other & Other & 16,447 & 5.7\% \\
cc-by-nc-4.0 & CC & 8,318 & 2.9\% \\
llama2 & ML & 6,158 & 2.1\% \\
unknown & Unknown & 4,419 & 1.5\% \\
cc-by-4.0 & CC & 4,269 & 1.5\% \\
llama3 & ML & 3,335 & 1.2\% \\ \hline
\end{tabular}%

\vspace{0.3cm}
\caption{Top 10 licenses for datasets.}
\label{tab:top_licenses_datasets}
\begin{tabular}{lcrr}
\multicolumn{4}{c}{\textbf{Datasets}} \\ \hline
\textbf{License} & \textbf{Type} & \textbf{Count} & \textbf{\begin{tabular}[c]{@{}r@{}}Percent \\ (of licensed)\end{tabular}} \\ \hline
mit & OSS & 13,501 & 27.9\% \\
apache-2.0 & OSS & 12,372 & 25.6\% \\
openrail & ML & 5,501 & 11.4\% \\
cc-by-4.0 & CC & 2,673 & 5.5\% \\
unknown & Unknown & 2,194 & 4.5\% \\
other & Other & 1,837 & 3.8\% \\
cc-by-sa-4.0 & CC & 1,256 & 2.6\% \\
cc-by-nc-4.0 & CC & 1,014 & 2.1\% \\
cc-by-nc-sa-4.0 & CC & 946 & 2.0\% \\
cc0-1.0 & CC & 799 & 1.7\% \\ \hline
\end{tabular}%
\end{table}

There were 71 unique declared licenses for models and 70 for datasets. %
\Cref{tab:top_licenses_datasets,tab:top_licenses_models} provide an overview of the ten most frequently observed licenses. %

We classify licenses by the categories defined above. %
A breakdown of the prevalence of these classes can be found in \Cref{tab:class_prevalence}. Both models (62.5\%) and datasets (58\%) were most often licensed under approved OSS licenses. %
This is particularly interesting with respect to datasets, as we would expect to see primarily CC or Data licenses, as datasets are not ``software'' in a strict sense of the word. %
This may reflect that Hugging Face developers view models and datasets more similarly to programs than to data or that developers rely on OSS licenses because they are already familiar with their terms. However, \add{as prior work indicates \cite{duan2024modelgo},} the relationships between these preexisting licenses and these new types of components in the ML supply chain, as well as the \add{legal and other} implications of licensing such components with such licenses, are currently unknown. Further research is needed to understand the reasoning behind these licensing choices, as well as to understand the interactions between many of these licenses and ML components.

\begin{table}[]
\caption{Prevalence of License Classes}
\label{tab:class_prevalence}
\begin{tabular}{lrrclrr}

\multicolumn{3}{c}{\textbf{Models}} & \textbf{} & \multicolumn{3}{c}{\textbf{Datasets}} \\
\cline{1-3} \cline{5-7}
\textbf{License Class} & \textbf{Count} & \textbf{Percent} & \textbf{} & \textbf{License Class} & \textbf{Count} & \textbf{Percent} \\ \cline{1-3} \cline{5-7} 
OSS & 179,421 & 62.5\% &  & OSS & 28,047 & 58.0\% \\
ML & 66,540 & 23.2\% &  & CC & 9,151 & 18.9\% \\
CC & 19,856 & 6.9\% &  & ML & 6,481 & 13.4\% \\
Other & 16,447 & 5.7\% &  & Unknown & 2,194 & 4.5\% \\
Unknown & 4,419 & 1.5\% &  & Other & 1,837 & 3.8\% \\
Data & 326 & 0.1\% &  & Data & 559 & 1.2\% \\
Combination & 134 & 0.05\% &  & Combination & 87 & 0.2\% \\ \cline{1-3} \cline{5-7} 
Total Licensed & 287,143 & 100.0\% &  & Total Licensed & 48,356 & 100.0\% \\ \cline{1-3} \cline{5-7} 
\end{tabular}%
\end{table}

Both models and datasets are distributed with (i) OSS licenses specific to software (both restrictive and permissive), (ii) Creative Commons (CC) licenses, which are not software-specific yet are being used for software (e.g., Stack Overflow's adoption of CC licenses), and ML-specific licenses. For models, our findings are consistent with those of Pepe \etal  \cite{pepe2024hugging}, %
showing that permissive OSS licenses such as \license{Apache-2.0} and \license{MIT} are popular in this space. This is not surprising, as such licenses are also among the most popular for open-source projects~\cite{top_licenses,vendome2015large} 
and allow commercial/closed-source exploitation.  We note that, in general, the top licenses adopted on Hugging Face %
are traditionally understood to be more permissive, even if their exact application in this novel context is not fully understood.

We also note the adoption (for models, but especially for datasets) of different variants of the CC license, including a permissive license (\license{CC-BY}), restrictive licenses (\license{CC-BY-SA} and \license{CC-BY-NC-SA}), and licenses limiting non-commercial use only (\license{CC-BY-NC}).%

Besides OSS licenses, we see a large proportion of different kinds of ML-specific licenses. These include both licenses originating from open-source initiatives (\eg \license{openrail} and \license{creativeml-openrail-m}), and licenses originating from companies (\eg Meta's \license{llama2} and \license{llama3} licenses). 
Such licenses tend to have terms that differ from typical redistribution terms established by software licenses such as, for example, terms that %
require responsible model usage or prohibit adoption for harmful or unethical uses or uses that do not take models' limitations into account. Likewise, the \license{llama2} and \license{llama3} licenses %
impose limitations related to commercial use,  use in products or services having %
more than 700 million monthly active users, and use to train other models, unless such models are redistributed, under the same licenses, as derivative works of \license{llama} models. %

 Lastly, we note the increased prevalence of Hugging Face's \other license, %
 which was ranked 7th in a study conducted by Pepe \etal \cite{pepe2024hugging}, but has since moved to the 5th spot since.  %
 In fact, 3.8\% of datasets and 5.7\% of models are licensed with some \other license. %

 \subsection{Exploration of the \other License}
\label{ssec:other_license}

In our entire dataset, 5.7\% of licensed models were labeled as being  under some \other license. This makes the \other category the fifth most common licensing choice for models on Hugging Face, representing over sixteen thousand models.  %
Non-standard, custom/vanity licenses can complicate compliance processes for organizations and individuals, given that the documents no longer represent a known quantity~\cite{wintersgill2024law}. To assess the nature and potential impact of these licenses, we investigate these \other licenses in further detail.

\subsubsection{Manual analysis of models under the \other license}
\label{sssec:manual_other_license}

We examine the licenses of the top 25 \other-licended models, sorted by downloads%
, as well as those of 25 additional, randomly sampled \other-licensed models from the total population. %
Our assumption is that the popularity of models in the high-downloads group indicates a tendency toward high-quality documentation practices, allowing us to investigate a detailed and meaningful sample of projects licensed in this way while maintaining a reasonable sample size for manual analysis. We also select a random sample in an attempt to determine if the same licensing trends exist across the ecosystem, even for less popular models. One author manually and systematically reviewed the model pages of these 50 models to identify more specific licensing information.  
Cumulatively, these 50 models represent 13,945,199 downloads (0.8\% of all model downloads). The results can be seen in Tables \ref{tab:other_license_finds} and \ref{tab:other_license_finds_random}.  (One of the models randomly sampled was no longer available on Hugging Face for analysis.) %

\begin{table}[]
\caption{``Other'' license information for Top 25 ``Other''-licensed models }
\label{tab:other_license_finds}
\resizebox{\columnwidth}{!}{
\begin{tabular}{|lll|}
\hline
\textbf{Model Name}                                 & \textbf{Resolved License}                                                   & \textbf{Find Location}                                                     \\ \hline
facebook/opt-125m                                   & OPT-175B LICENSE AGREEMENT                                                  & LICENSE.md                                                                 \\
stabilityai/sdxl-turbo                              & sai-nc-community                                                            & LICENSE.md                                                                 \\
coqui/XTTS-v2                                       & coqui-public-model-license                                                  & LICENSE.txt                                                                \\
facebook/mask2former-swin-base-coco-panoptic        & None Available*                                                             & None Available                                                             \\
briaai/RMBG-1.4                                     & BRIA 2.0 on HuggingFace Model License Agreement                             & \begin{tabular}[c]{@{}l@{}}External Link\\ (connected to tag)\end{tabular} \\
facebook/mask2former-swin-large-cityscapes-semantic & None Available*                                                             & None Available                                                             \\
apple/DFN5B-CLIP-ViT-H-14-378                       & apple-sample-code-license                                                   & LICENSE                                                                    \\
facebook/opt-1.3b                                   & OPT-175B LICENSE AGREEMENT                                                  & LICENSE.md                                                                 \\
facebook/mask2former-swin-large-ade-semantic        & None Available*                                                             & None Available                                                             \\
Lykon/DreamShaper                                   & \begin{tabular}[c]{@{}l@{}}CreativeML Open RAIL-M\\ Addendum*\end{tabular} & \begin{tabular}[c]{@{}l@{}}External Link\\ (in model card)\end{tabular}    \\
Salesforce/instructblip-vicuna-7b                   & None Available*                                                           & None Available                                                             \\
SanctumAI/Meta-Llama-3-8B-Instruct-GGUF             & llama3                                                                      & LICENSE                                                                    \\
cagliostrolab/animagine-xl-3.1                      & faipl-1.0-sd                                                                & \begin{tabular}[c]{@{}l@{}}External Link\\ (connected to tag)\end{tabular} \\
stabilityai/stable-video-diffusion-img2vid-xt       & stable-video-diffusion-community                                            & LICENSE.md                                                                 \\
apple/mobilevit-small                               & apple-sample-code-license                                                   & In model card                                                              \\
jartine/gemma-2-27b-it-llamafile                    & Gemma License                                                               & In model card*                                                          \\
facebook/opt-350m                                   & OPT-175B LICENSE AGREEMENT                                                  & LICENSE.md                                                                 \\
Qwen/Qwen2-72B-Instruct-AWQ                         & tongyi-qianwen                                                              & LICENSE                                                                    \\
Mozilla/Meta-Llama-3-70B-Instruct-llamafile         & Llama3                                                                      & In tags*                                                              \\
digiplay/AbsoluteReality\_v1.8.1                    & Absolute Reality License*                                             & \begin{tabular}[c]{@{}l@{}}External Link\\ (in model card)\end{tabular}    \\
qwp4w3hyb/Qwen2-72B-Instruct-iMat-GGUF              & tongyi-qianwen                                                              & In tags*                                                               \\
google/mobilenet\_v1\_0.75\_192                     & None Available                                                              & None Available                                                             \\
Qwen/Qwen2-72B-Instruct                             & tongyi-qianwen                                                              & LICENSE                                                                    \\
mradermacher/QuartetAnemoi-70B-t0.0001-i1-GGUF      & None Available                                                              & None Available                                                             \\
SanctumAI/Codestral-22B-v0.1-GGUF                   & mnpl                                                                        & In tags*          \\ \hline                                                    
\end{tabular}
}
{
\\ \footnotesize * Information was ultimately inconsistent, unavailable, or changed since initial mining.
}
\end{table}

\begin{table}[]
\caption{``Other'' license information for random sample of ``Other''-licensed models }
\label{tab:other_license_finds_random}
\resizebox{\columnwidth}{!}{
\begin{tabular}{|lll|}
\hline
Model Name                                                            & License                 & Location                         \\ \hline
Shalie/EliraPendoraPonyXL                                             & faipl-1.0-sd            & External Link (connected to tag) \\
sukhoi37/sp\_weshi                                                    & None Available*         & None Available                   \\
Bobross69/Usagiyam                                                    & None Available*         & None Available                   \\
baratk/barat                                                          & Empty file              & LICENSE file                     \\
styalai/competition-math-phinetune-v1                                 & None Available*        & None Available                   \\
divyanshu074/Text\_Beat                                               & None Available*         & None Available                   \\
LoneStriker/WhiteRabbitNeo-33B-v1-3.0bpw-h6-exl2                      & deepseek                & External Link (connected to tag) \\
jiuhai/llama3-ift                                                     & REMOVED*              & REMOVED*                       \\
KnutJaegersberg/Deacon-34B-qlora                                      & yi-license              & LICENSE file                     \\
second-state/Qwen1.5-1.8B-Chat-GGUF                                   & tongyi-qianwen-research & LICENSE file                     \\
psychicfire/sample                                                    & None Available*         & None Available                   \\
Workaholic/fake-pvc-style                                             & None Available*         & None Available                   \\
azi111/dolphin-2\_2-yi-34b-465bpw-h8-exl2-cnen                        & Empty file              & LICENSE file                     \\
dranger003/deepseek-coder-33b-instruct-iMat.GGUF                      & deepseek                & External Link (connected to tag) \\
optimai/porne-0                                                       & None Available*         & None Available                   \\
cortecs/Meta-Llama-3-70B-Instruct-GPTQ-8b                             & None Available*      & None Available                   \\
TheBloke/MergeMonster-13B-20231124-GPTQ                               & Other / llama2          & Model card                       \\
yzhuang/Meta-Llama-3-8B-Instruct\_fictional\_Chinese\_v3              & None Available*         & None Available                   \\
fasterinnerlooper/stable-code-3b                                      & None Available*         & None Available                   \\
DevQuasar/Qwen2-72B-Instruct-GGUF                                     & tongyi-qianwen          & External Link (connected to tag) \\
titan087/Llama3-70B-ShiningValiant2-exl2-4b                           & llama3                  & External Link (connected to tag) \\
nulltella/phi-1\_5-alcapa-51k-instruct                                & None Available*         & None Available                   \\
Holarissun/mar13\_gemma2b\_aisft\_gsm8k\_rand\_alpha0.9995-subset7000 & None Available*         & None Available                   \\
raincandy-u/Quark-464M-v0.1.alpha                                     & tongyi-qianwen-research & External Link (connected to tag) \\
TheBloke/Tess-34B-v1.4-GGUF                                           & yi-34b                  & External Link (connected to tag) \\ \hline
\end{tabular}
}
{
\\ \footnotesize * Information was ultimately inconsistent, unavailable, or changed since initial mining.
}
\end{table}

\subsubsection{Top 25 \other licensed models}

For the top 25 models, additional licensing information was most often found in a LICENSE file that was associated with the repository, but four (16\%) linked to an external page that included more licensing information.  There were four cases (16\%), including three Facebook models%
, where a specific license could not be determined, but one could be inferred from a related GitHub repository, which often was not directly linked to the model card.  We also observed a case, \model{Lykon/DreamShaper}, where the license that was specified in the model card did not match the external reference also provided.  

Determining the licenses for some models required looking beyond the information provided on Hugging Face itself. For example, the README for the model \model{digiplay/AbsoluteReality\_v1.8.1} included a link to a model page on an external site (Civitai.com).  On this page, the license is reported as \license{CreativeML Open RAIL-M Addendum}.  Originally, clicking the link associated with the license would take the user to a page for the \license{Absolute Reality License} (a confusing license mismatch), but in the time since, the link has been corrected to point to the full text of the \license{CreativeML Open RAIL-M Addendum}.%

We can also see there were three instances where a model reported an \other license, but the actual license specified was in Hugging Face’s recognized license list (either \license{llama3} or \license{gemma}).  This could be the result of the model owner uploading their model before the license was officially recognized or a general unfamiliarity on the uploader’s part with the available license options.  In either case, this particular error further underscores the manual and error-prone documentation processes on Hugging Face.

Somewhat surprisingly, even after systematic manual documentation searches (including web searches for associated GitHub repositories), there were two models in the top 25 \other licensed models (8\%) where the specific license could not be identified.  From a practical standpoint, these models might as well not be licensed at all, since users cannot comply with unavailable terms and conditions.

Interestingly, two of the models that had no licensing information included in their documentation, \model{Salesforce/instructblip-vicuna-7b} and \model{google/mobilenet\_v1\_0.75\_192}, actually had model cards produced by the Hugging Face team.

\subsubsection{Random sample of 25 Other-licensed models}

Next, we examined an additional random sample of 25 models that were reportedly under an \other' license.  This would allow us to determine if the licensing practices followed by the top 25 are representative of the ecosystem.  As one might expect, we found that identifying the licenses for many of these models was difficult.  Ten models (40\%) had model cards that were void of all text. One additional model had some information in its model card but none that pertained to licensing. Two more models had LICENSE files in their repositories, but the files were empty. Another model, \model{Meta-Llama-3-70B-Instruct-GPTQ-8b}, provided no further licensing information, but given its name, one might infer it was under the \license{Llama3} license.

Some of the models, however, did include more specific licensing information. Six (24\%) had links to external references included in their metadata. One model, while labeled as licensed under \other, was actually under the \license{Llama3} license. Two models had LICENSE files that provided additional license information.

In the text of the model card for \model{TheBloke/MergeMonster-13B-20231124-GPTQ}, we observed the model owner grappling with certain ambiguities and uncertainties surrounding the \other license:
``The creator of the source model has listed its license as other, and this quantization has therefore used that same license. As this model is based on Llama 2, it is also subject to the Meta Llama 2 license terms, and the license files for that are additionally included. It should therefore be considered as being claimed to be licensed under both licenses. I contacted Hugging Face for clarification on dual licensing but they do not yet have an official position. Should this change, or should Meta provide any feedback on this situation, I will update this section accordingly. In the meantime, any questions regarding licensing, and in particular how these two licenses might interact, should be directed to the original model repository.'' 
Here, we see that properly maintaining license compliance can be a difficult and confusing task, particularly when all relevant information is not made readily available.  It is also important to note that since the original model being quantized was itself a derivative of a \model{Llama2} model, this model uploader has restored the \license{Llama2} license to remain compliant.

\subsubsection{Automated analysis of models under the Other license}
\label{rq2:other-analysis}

Having determined from the manual review of 50 models reportedly under the \other license how and where to find more specific licensing information, we next moved to an automated approach that could provide insights about the most common \other licenses in the ecosystem.  We note that since this analysis required access to more data than was originally pulled for the rest of the dataset described in this paper, we conducted a separate smaller mining on February 1, 2025.  Specifically, we pulled the top ten thousand models (sorted by downloads) that reported they were under the \other license.  We acknowledge that the results of this subsequent mining may not perfectly align with the dataset presented in the rest of this work, but nonetheless, we believe it is worthwhile to explore the phenomenon of the \other license in more depth.

\begin{table}[]
\caption{\other licenses reported in \cardData attribute in Top 10,000 \other-licensed models %
}
\label{tab:cardData_licenses}
\begin{tabular}{lll}
\hline
\textbf{License Name}             & \textbf{Count} & \textbf{Percentage} \\ \hline
faipl-1.0-sd                      & 2346           & 23\%                \\
flux-1-dev-non-commercial-license & 1172           & 12\%                \\
llama3                            & 269            & 3\%                 \\
deepseek                          & 206            & 2\%                 \\
tongyi-qianwen                    & 201            & 2\% \\ \hline                
\end{tabular}
\end{table}

\Cref{tab:cardData_licenses} shows the top 5 specific licenses that were reported in the \cardData attribute.  Over a third of models (36\%) did not provide additional licensing information in their \cardData.  Three percent of models are actually under the \license{llama3} license, which is in the recognized license list on Hugging Face \cite{hugging_face_licenses}.  As mentioned before, this suggests that either these models were uploaded before \license{llama3} was added to the official list and have not been updated or that the model owners are unfamiliar with the licenses available in the official list. In either case, feasible tooling solutions can be readily imagined to resolve these problems. %

There is a further lack of standardization in the licenses reported as \other.  For example, 206 models report \license{deepseek} and an additional 85 report the \license{deepseek-license}, though these are presumably the same license.  As previously mentioned, 269 models erroneously report \license{llama3} as their \other license, but an additional 27 report being under the \license{llama-3} license.  We see 110 models under the \license{gemma-terms-of-use}, even though the proper identifier for this model is just \license{gemma}~\cite{hugging_face_licenses}.  All of this further underscores the problems inherent in providing developers with a free-text field instead of assisted tool support that suggests known licenses.

\subsection{Parent/child license differences} %
\label{sec:base-to-deriv-variations} %

\begin{table*}[]
\caption{Parent/child license difference categories and their frequencies} %
\label{tab:license_change_frequencies}
\resizebox{\columnwidth}{!}{
\begin{tabular}{lllrr}
\hline
\textbf{Difference Type} & \textbf{Description}                                                                                                                                                                                            & \textbf{Example}                                           & \textbf{Count}  & \textbf{Percentage} \\ \hline
No Overlap               & \begin{tabular}[c]{@{}l@{}}Both the parent and child are licensed, but\\ there is no overlap in their licensing\end{tabular}                                                                                & cc-by-nc-4.0 $\rightarrow$ apache-2.0                      & 36,441          & 54.8\%              \\
Licensing Dropped        & \begin{tabular}[c]{@{}l@{}}The parent is under one or more licenses,\\ but the child has no specified license\end{tabular}                                                                                      & apache-2.0 $\rightarrow$ UNDECLARED                        & 19,435          & 29.2\%              \\
Licensing Added          & \begin{tabular}[c]{@{}l@{}}The child is under one or more licenses, but\\ the parent has no specified license\end{tabular}                                                                                      & UNDECLARED $\rightarrow$ apache-2.0                        & 10,548          & 15.9\%              \\
License(s) Added         & \begin{tabular}[c]{@{}l@{}}In a multi-licensing scenario, the child retains\\ all of its parent's licenses and adds one or more\\ additional licenses.\end{tabular}                                             & apache-2.0 $\rightarrow$ apache-2.0 $\vert$ mit            & 28              & 0.04\%              \\
License(s) Removed       & \begin{tabular}[c]{@{}l@{}}In a multi-licensing scenario, the child is no\\ longer under one or more licenses that apply\\ to its parent, but is still licensed and adds no\\ additional licenses.\end{tabular} & apache-2.0 $\vert$ bsd-3-clause $\rightarrow$ bsd-3-clause & 8               & 0.01\%              \\ \hline
\textbf{Total Changes}   & \textbf{-}                                                                                                                                                                                                      & -                                                          & \textbf{66,460} & \textbf{100.0\%}    \\ \hline
\end{tabular}%
}
\end{table*}

\begin{table}[]
\caption{Most common parent/child license differences}
\label{tab:license_changes}
\begin{tabular}{llrr}
\hline
\textbf{Parent License} & \textbf{Child License} & \textbf{Count }& \textbf{Percentage }\\ \hline
cc-by-nc-4.0 & apache-2.0 & 11,001 & 30.19\% \\
apache-2.0 & cc-by-nc-4.0 & 7,427 & 20.38\% \\
cc-by-4.0 & cc-by-nc-4.0 & 2,626 & 7.21\% \\
other & llama3 & 1,720 & 4.72\% \\
cc-by-nc-4.0 & cc-by-4.0 & 1,444 & 3.96\% \\
other & apache-2.0 & 1348 & 3.70\% \\
apache-2.0 & mit & 1,241 & 3.41\% \\
cc-by-nc-nd-4.0 & cc-by-nc-4.0 & 1,180 & 3.24\% \\
llama3 & other & 1,083 & 2.97\% \\
apache-2.0 & cc-by-4.0 & 782 & 2.15\%\\ \hline
\end{tabular}%
\end{table}

\begin{figure}[t]
\centering
\includegraphics[width=0.4\linewidth]{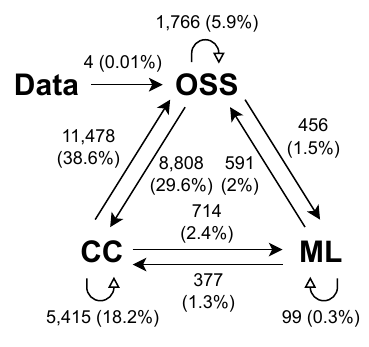}
\caption{Parent/child license differences across categories (Total differences: 29,708)}%
\label{fig:changes_across_classes}
\end{figure}

Our dataset contains 274,104 distinct parent/child relationships.  We observed 66,460 instances (24\%) where the licensing of a child model (\ie derivative model) was different from that of its parent (\ie base model). \Cref{tab:license_change_frequencies} provides an overview of the types of differences observed. In nearly a third of cases, child models specified no licensing information despite information being available for their parent(s), leading to a potential compliance issue. We also observe that once license information has been dropped, it is unlikely to be restored by later links in the chain, with license restoration behavior observed in only 15.9\% of instances.  A complete shift in licensing between parent and child was seen in 54.8\% of cases (\ie no licenses were shared between the two).  Of these shifts, 18.5\% involved the \unknown or \other license category.  \Cref{tab:license_changes} shows the top ten such shifts, and \Cref{fig:changes_across_classes} provides more information on how licensing decisions shifted across categories between parents and their children (\eg a difference from \license{cc-by-4.0} to \license{MIT}), as well as how licensing decisions shifted within categories (\eg a difference from \license{MIT} to \license{Apache-2.0})%
While ML-specific licenses seem a good fit for models, model owners may be opting for licenses they are more familiar and comfortable with (OSS for software developers and CC for data scientists). We observe that most shifts occur between the OSS and CC categories (68.3\%), perhaps suggesting some tension between the licensing preferences of those with data science and software development backgrounds. Additionally, depending on the specifics of the situation and the licenses involved, these shifts can also potentially introduce license incompatibilities, such as dropping non-commercial requirements or failing to apply the requirements of a copyleft license.

\subsection{Dataset license and model license combinations} 

\begin{table}[]
\caption{Model license and dataset license pairs}
\label{tab:dataset_model_licenses}
\begin{tabular}{llrr}
\hline
\textbf{Model License} & \textbf{Dataset License} & \textbf{Count} & \multicolumn{1}{l}{\textbf{Percentage}} \\ \hline
apache-2.0 & other & 8196 & 18.9\% \\
apache-2.0 & cc0-1.0 & 4621 & 10.6\% \\
mit & mit & 4479 & 10.3\% \\
apache-2.0 & apache-2.0 & 3916 & 9.0\% \\
apache-2.0 & unknown & 2993 & 6.9\% \\
apache-2.0 & mit & 2734 & 6.3\% \\
mit & other & 2290 & 5.3\% \\
apache-2.0 & cc-by-4.0 & 2076 & 4.8\% \\
apache-2.0 & cc-by-sa-4.0 & 1934 & 4.5\% \\
mit & apache-2.0 & 1739 & 4.0\% \\ \hline
\end{tabular}%
\end{table}

The licenses of datasets that are used during training can have an impact on the licensing decisions for the final ML model.  Here we consider the pairings of dataset licenses and the resultant model licenses.  There are 623 distinct model/dataset license combinations within our dataset across 43,455 such pairings; however, since we (and Hugging Face) aggregated all \other licenses under one label, these numbers likely overestimate the consistency in the space. %
(The top ten most frequent can be found in \Cref{tab:dataset_model_licenses}.)  %
The license of the model exactly matches the license of at least one dataset it was trained on in only 41 of the 623 combinations. %
The most common combination is a model licensed under the \license{apache-2.0} license and a dataset under a custom or \other license, with 11,731 (27\%) pairs involving a dataset under a custom or \other license. This can make using these datasets and models problematic since retrieving and evaluating the licensing terms can be difficult and time-consuming. %

\subsection{Additional licensing findings}

\subsubsection{Multi-licensing.}
\label{sec:multilicensing}
A few models (188) %
and datasets (104) were released under more than one license. Multi-licensing is an existing phenomenon in the world of open-source software but is made more complex in the ML model context by the presence of novel license combinations that are not yet well understood.  Examples include the 48 models we observed multi-licensed under \license{apache-2.0} and \license{cc-by-nc-4.0} as well as models under both OSS and ML licenses, such as \license{MIT} and \license{OpenRAIL}. We observe examples of such multilicensing in the most declared datasets in \Cref{tab:top_declared_datasets}. %
We also observed datasets that were released under as many as six distinct licenses. %
The Hugging Face metadata provides no way of determining the relationship among these licenses, which may be an AND or an OR, or even indicate, as observed in \Cref{sec:case-model-dataset-licensing,sec:case-dataset-licensing}, that portions of the dataset are under different licenses. %
Regardless, this distinction is critical to determining compliance.  For example, if a user who uploaded a \license{llama2} derivative model chose to make it available with an OR relationship between the \license{apache-2.0} and \license{llama2} licenses, this would contradict the exclusive \license{llama2} licensing of its base model, resulting in a violation that could propagate to models further down the chain. %
While we observed only a few instances of multi-licensing, such activity motivates the need to understand the interplay between the various license classes and for Hugging Face to supply a standardized way to specify the intended relationship between licenses.

As a case study, we consider the most common example of multi-licensing found across models: a combination of the \license{apache-2.0} and \license{cc-by-nc-4.0} licenses. Depending on the relationship between these two licenses and the reasoning for including them, the combination could represent potential compliance challenges. To explore this, we conduct a manual analysis of \model{mradermacher/ NSFW\_DPO\_Noromaid-7b-Mistral-7B-Instruct-v0.1-GGUF}, the model under this licensing scheme with the most downloads.  This model is a quantization of \model{MaziyarPanahi/
NSFW\_DPO\_Noromaid-7b-Mistral-7B-Instruct-v0.1}, which is also dual licensed in the same way.  The model card for\model{MaziyarPanahi/NSFW\_DPO\_Noromaid-7b-Mistral-7B-Instruct-v0.1} describes it as a merge of two additional models.  The first, \model{mistralai/Mistral-7B-Instruct-v0.1}, is under the \license{apache-2.0} license, and the second, \model{athirdpath/NSFW\_DPO\_Noromaid-7b}, is under the \license{cc-by-nc-4.0} license.  Following the supply chain back further, we observe that the dataset that \model{athirdpath/
NSFW\_DPO\_Noromaid-7b} was trained on was also licensed under \license{cc-by-nc-4.0}.  \model{mistralai/Mistral-7B-Instruct-v0.1} is a finetune of \model{mistralai/Mistral-7B-v0.1}, but there is no record of the data used for training.  From all of this we see that the licensing relationship in this multi-licensing example is likely intended to be an AND (\ie that the model is released under the terms of both the \license{apache-2.0} and \license{cc-by-nc-4.0} licenses).  However, if other model trainers were to interpret the relationship as an OR (\ie the model is licensed such that creators of derivative works can freely choose which licensing terms, either the \license{apache-2.0} or \license{cc-by-nc-4.0} license, to follow), it is possible that the non-commercial limitation imposed by the \license{cc-by-nc-4.0} license could be dropped and non-compliance could be introduced.  We also note that here, instead of finding a single suitable license for \model{MaziyarPanahi/NSFW\_DPO\_Noromaid-7b-Mistral-7B-Instruct-v0.1}, the model owner opted to just include the licenses of both dependencies.  While not observed in this case, it is possible that this strategy could fail in certain instances where dependency licenses conflict.

\subsubsection{Naming requirements.} 
\label{sec:naming_requirements}
Some licenses, like \license{Llama3}, impose a %
generational limitation \cite{oss_book} that specifies naming requirements for derivative works.  According to the license, models that are built on a model under this license must have names prefixed with ``llama3'' \cite{llama3_license}.  In our dataset, there are 3,335 models licensed under \license{llama3}. However, of these, only 473 (14.2\%) models correctly have names beginning with ``llama3'' as required by the license.  Only 768 (23\%) models even contain the term ``llama3'' in their name.  Indeed, just 2,421 (72.6\%) models contain the term ``llama'' at all, and 384 (11.5\%) only contain the string ``l3.'' %
The problem is likely even worse in practice, as these statistics only include models that documented that they were under the \license{llama3} license.  %

\subsubsection{Attribution.}
As was observed during the manual model card analysis described in \Cref{sec:license_extraction}, attribution appears to be important to model owners regardless of licensing status. Rather than requesting or requiring attribution through a license, models without a license often asked for some form of attribution (29/84), typically in the form of citations. %
In fact, 83,628 models (11.1\%) mined using the Hugging Face API included a link to a paper on \textit{arxiv} \cite{arxiv} in their tags, thereby providing a means of citing the work behind the model.  In our manual analysis of model cards, we observed that the model owner was typically also affiliated with (\eg an author on) the linked research paper. %
This desire for recognition is reflected by the top licenses in the space, namely \license{MIT} and \license{Apache-2.0}, each of which specifies attribution requirements. (A similar historical example is the evolution of Creative Commons licenses in the early 2000s, where the overwhelming preference for attribution led to its becoming a default term \cite{standard_attribution}.) Given the research community's adoption of Hugging Face, %
it may be worthwhile to add dedicated metadata fields to facilitate the standardization of citation information. 

\subsection{Real World Potential Incompatibilities}
Here, we go beyond the parent/child license differences explored in \Cref{sec:base-to-deriv-variations} and examine some observed potential incompatibilities. %
We note that the list of compatibility issues is not exhaustive, since there are many unresolved legal questions concerning the application of certain terms.  Here we report the two most frequently observed issues in our dataset. While other issues were present, each only represented a few dozen examples in the entire ecosystem, so we exclude them from this section. We also exclude from this section any mappings that involve \other, \unknown, or undeclared licenses. 

\subsubsection{Dropping of non-commercial requirements}

30.19\% of parent/child license differences involved a parent under \license{cc-by-nc-4.0} and a child licensed under \license{apache-2.0}.  An additional 3.96\% involved a difference between the parent license \license{cc-by-nc-4.0} and the child license \license{cc-by-4.0}.  In both cases, the ``nc'' in the parent license indicates that the work can not be used for commercial purposes.  The \license{apache-2.0} license provides no such limitation, thus introducing a potential conflict.  
A change in requirements of such a nature could potentially lead to a license violation. Consider a situation in which a developer releases a component under \license{cc-by-nc-4.0}, but later, a different developer fine-tunes or packages it under the \license{mit} license. A user might seek to use this \license{mit}-licensed version of this component in a commercial product, thinking the \license{mit} license applies, though in the process they violate the non-commercial requirement of the \license{cc-by-nc-4.0} license that the component was originally made available under.
In this sense, we may observe something akin to unintentional (or perhaps intentional) license laundering \cite{longpre2023data}. %
    
\subsubsection{Violation of no derivatives term}

Of the total parent/child license differences, 3.24\% featured the parent license \license{cc-by-nc-nd-4.0} and the child license \license{cc-by-nc-4.0}.  The only difference between the two being the dropped ``nd'' condition, which indicates that no derivatives based on the licensed work are permitted.  This means that if we consider a fine-tuned model to be a derivative of its base model, then all 1,180 of these instances represent an example of noncompliance.%

\subsection{Summary}

Overall, licensing on Hugging Face presents a number of unique challenges. While traditional open source software licenses remain popular among ML components, the introduction of ML- and data-specific licenses raise novel questions regarding the compatibility of different licenses, and, as such, different components. The \unknown and \other licenses complicate the matter of compliance further by obscuring licensing information. To illustrate the specific difficulties that may arise, we gathered classes of potential licensing issues currently present in models on Hugging Face.

\section{Case study: top models}
\label{ssec:case_study_models}

To better understand and illustrate the supply chain structure and challenges described in this section, we present a detailed analysis of the top 100 most downloaded models and the top 100 most liked models from our dataset. As described in \Cref{sssec:manual_other_license}, we suspect that model downloads can be used as a proxy for documentation quality. %

\subsection{Sample construction}
Among the 100 top downloaded and 100 top liked models, there were 187 unique models. The 13 overlapping models were mostly uploaded by well-known technology companies, including OpenAI, Meta, StabilityAI, Google, and Microsoft. The only exception was the \model{all-MiniLM-L6-v2} model uploaded by sentence-transformers. %
These 187 models collectively yielded 1,465,884,673 downloads, or 79.7\% of all 1,838,905,161 downloads from the entire dataset of 760,460 models.

Forty-five (24\%) are \textsc{text-generation} models, 28 (15\%) are \textsc{text-to-image} models, and 17 (9\%) are \textsc{text-classification} models.  Other common model types included \textsc{fill-mask} (7\%), \textsc{speech recognition} (6\%), and \textsc{sentence-similarity} (5\%). We also note that 21 (11\%) models did not specify the pipeline (\ie a high-level abstraction that simplifies using models for common tasks, roughly corresponding to the model's intended task~\cite{pipelines}), revealing documentation shortcomings even for some popular models. %

By far, the most common libraries (\ie open-source tools available through Hugging Face that support ML training~\cite{libraries}) used were \library{transformers} (121) and \library{sentence-transformers} (11).  %
The \library{diffusers} library was
the next most common with 28 instances. Twelve models (6.4\%) did not specify in their documentation any libraries.

Ninety-three unique model owners are represented in the top 187 models, including organizations like StabilityAI, Google, Meta, Microsoft, and OpenAI.  In fact, 91 of the 187 top models (48.6\%) are owned by just 15 model uploaders (without combining owners such as meta-llama, facebook, and facebookai into one entity), suggesting that a small handful of players dominate the landscape. Sixty-three model uploaders (67.7\%) have only one model in the top 187 models, suggesting long-tail behavior similar to that observed with respect to the dataset as a whole in \Cref{sssection:model_ownership}.

Of the 187 models, 148 models (79\%) are relied on by at least one other model available on Hugging Face (\ie they are base models).  When considering the supply chain graph discussed in \Cref{sec:rq1}, we see that the 148 base models collectively have 43,976 outgoing edges, with a mean out-degree of 235 and a standard deviation of 665.  Twenty-one models have an out-degree of one, and twelve have an out-degree of two.  In total, 43,874 distinct models depend on this set of 148 models, further demonstrating their relevance to the ecosystem.

\subsection{Documented base models}

Four models (2.1\%) in the top 187 declare a base model of their own (\ie they depend on another model) .%
Given that many of the models in our set are foundation models, we would  not expect them to be built on other publicly available models. 
Only three distinct base models were declared in the set, two of which were contained in the top 187 models themselves (\model{mistralai/Mistral-7B-v0.1} and \model{xlm-roberta-base}). Despite the dearth of base model declarations, 8 models showed evidence of quantization (\ie completed the ``quantization\_config'' or ``quantized\_by'' fields in its metadata or included a file containing the substring ``quant'' in the repository) and were confirmed to be quantizations through manual analysis. However, none of these quantized models declared a base model in the appropriate metadata field (\eg, \cardData/basemodel), yet by virtue of being quantizations, these models must depend on a base model. This example illustrates the ongoing need for more thorough documentation and declaration of base models on Hugging Face.

In general, the supply chain of models in the top 187 is shallow. Few models declared a base model, and those that did relied on a single foundation model with no further dependencies. There were no instances in which a popular model relied on a less popular one. There were two parent/child license differences, a switch from \license{openrail++} to \license{creativeml-openrail-m} and a switch from \license{apache-2.0} to \license{mit}, but neither of these are likely to result in compliance challenges, given the permissive natures of the licenses involved.%

\subsection{Model licensing analysis}

Next, we explore the presence of licensing information for models within our 187-model sample. Of that sample, only 156 provided any sort of licensing information, and of those, %
80 percent had licensing information in both their tags and in their \cardData field, 16\% had licensing information only in their tags, and no models had licensing information exclusively in their \cardData. %
This further reinforces that when considering completeness, the tags are the best source to mine.  However, while the tags seem to be the most complete source of licensing information with respect to determining what license is declared, the \cardData field, when completed properly, includes more specific information, such as external links to license files.
Twenty-six models (16\%) had a LICENSE file included in their repository.  Of these, twenty-two models (84.6\%) also had licensing information available in both their tags and \cardData fields, and four models reported licensing information exclusively through the LICENSE file. %

Conversely, there were twenty-four models (13\%) that did not include any licensing information in their tags, \cardData field, or a LICENSE file.  We were able to determine the license for only two of these models (1\%) 
through manual analysis.  %
Six models (25\%) included %
text in their documentation that provided terms, conditions, or guidance on derivative uses but which was not explicitly presented as licensing information. %
(One of these six models stated that the model was open-source, but provided no specifics.)  Eight of the twenty-four models (33\%) provided a %
BibTeX citation, suggesting that the model owners either want credit for their model or that they want to ensure that the original trainers receive credit. 

Lastly, we note that in three cases licensing information was found only in a seemingly related GitHub repository, but it was often unclear whether the license applied to the model weights, for the code that runs/loads them, or both. %

Ultimately, most model licenses in our sample are specified in the tags, \cardData, or in a LICENSE file. %
\Cref{tab:sample_license_info_loc} summarizes where licensing information was found for the models in this dataset. We note that examining the tags alone captures 96 percent of license declaration information, where such information exists. %

\begin{table}[]
\caption{Licensing information location in Top 187 models}%
\label{tab:sample_license_info_loc}
\begin{tabular}{lcc}
\hline
\textbf{Location}    & \textbf{Count} & \textbf{Percentage of licensed (n=165)} \\ \hline
Tags and \cardData    & 132            & 80\%                \\
Tags only            & 27             & 16\%                \\
LICENSE file only    & 4              & 2\%                 \\
Model card text only & 2              & 1\%      \\ 
No licensing information & 22 & - \\ \hline            
\end{tabular}
\end{table}

Of our 187-model sample, 165 models (88\%) declared some identifiable license.  An additional six models (3\%) included information on derivative uses, although this information was not presented as a formal license. %
Together, 91\% of models in the sample include some type of usage-related information, leaving 9\% of models in the sample without any licensing information. Given that the sample contains the most downloaded and liked models, this lack of information creates a significant problem for users, who risk violating copyright law as a result.

Discrepancies between information sources were rare, but did occur. There were no license discrepancies encountered between the tags and the \cardData fields.  However, when comparing the license information available in the tags with included LICENSE files, six potential discrepancies were observed. %
One model reported in its tags to be under the \license{creative-openrail-m}, but provided a completely custom LICENSE file.  There were four instances where models declared being licensed under \license{openrail++} in their tags, but their LICENSE files were for \license{CreativeML Open RAIL++-M}, which is a subset of \license{Open Rail++}.  %
Lastly, the model \model{stabilityai/stable-video-diffusion-img2vid-xt} was reportedly released under something called the stable-video-diffusion-community, but in reality, the LICENSE file was a verbatim copy of the STABILITY AI COMMUNITY LICENSE AGREEMENT (referred to as \license{sai-nc-community} elsewhere).\laura{Should the Stability AI Community License Agreement also be in small caps?} \nathan{The small caps are for the standardized names of the licenses}

Overall, we observed that the most popular models are likely to be permissively licensed.  Specifically 66 of the licensed models (40\%) were under the \license{apache-2.0} license, with an additional 35 models (21\%) made available under the \license{MIT} license.  The next largest group, with 21 models (13\%), were labeled with an \other license, explored in greater detail in \Cref{ssec:other_license}.  Thirteen models (8\%) were under the \license{creativeml-openrail-m} license, and six (4\%) were under \license{openrail++}.  Other licenses with at least two instances included \license{llama2} (5), \license{cc-by-nc-4.0} (5), \license{llama3} (3), \license{openrail} (2), \license{gemma} (2), \license{bsd-3-clause} (2), and \license{cc-by-4.0} (2). None of the most common models were declared to be under more than one license. %

Only one model in the top models was licensed as \unknown.  This appears to be an error, however, since there is a form at the top of the model card requesting the user to ``Acknowledge license to access the repository,'' clarifying that the model is under the \license{Falcon-180B TII} license. As such, the \other license option would likely have been more apt here.

Inversely, the \model{OpenAssistant/oasst-sft-6-llama-30b-xor} model is reportedly under an \other license, but no additional licensing information is provided, potentially making the \unknown option more apt.  Some legal guidance is provided however: ``Due to the license attached to LLaMA models by Meta AI it is not possible to directly distribute LLaMA-based models. Instead we provide XOR weights for the OA models.''

The other twenty models under an \other license represent issuance under custom and vanity licenses.  Four %
appear to be modifications of more popular licenses such as \license{CreativeML Open RAIL-M}.  The others are unique licenses presumably written by the model owner or their organization. %
This suggests that model trainers are unaware of existing licenses, do not want to invest time researching the terms of existing licenses, or believe that existing licenses do not cover all terms, conditions, and use cases envisioned by the model trainers.  Future work on the software engineering front will need to consider tooling solutions to help model trainers choose licenses that fit their specific needs.  If it is the case that current licenses are not up to the task, it may be necessary for legal professionals and researchers to participate in drafting more robust and applicable licenses.

\subsection{Documented datasets used}
\subsubsection{Overview}
\label{sssec:top_models_datasets_overview}
Only 56 (30\%) of models in the sample record at least one training dataset in their metadata. While large technology companies may conceal training sources to mitigate the potential of legal liability~\cite{wired-meta-piracy}, this is still a rate three times higher than the 10\% observed across the ecosystem at large (\Cref{sec:most-dependend-upon})%
. All 56 models recorded dataset information in both the tags and in the \cardData attribute. A total of 107 datasets were declared across the sample of 187 models, though we could only verify the presence of 96 of these on Hugging Face. (The remaining eleven resulted in a 404 error when searching on Hugging Face.) Thirty-three models reported relying on only a single dataset, but the average was 3.4 with a standard deviation of 5. Three models recorded relying on as many as 21 datasets.

Even this small set highlights several challenges in uniquely identifying datasets, such as three datasets, all belonging to the same account, that had their unique identifiers changed when the dataset owner updated their profile name from \user{ehartford} to \user{cognitivecomputations}. %
Two additional datasets from \user{embedding-data} also had their names changed. Without Hugging Face maintaining a mapping between the old and new human-readable IDs, it would be extremely difficult to correctly determine these dependencies. Other ambiguities arise from the unstandardized nature of some declarations. For example, it is unclear whether the reference to ``wikihow'' by some of the top models %
refers to the \model{wangwilliamyang/wikihow} dataset available on Hugging Face, some other dataset that compiles/aggregates information from wikiHow, or that the model was trained by scraping the wikiHow website. %
We explore the problems with identifying datasets further in \Cref{sec:missing_refs}.

\refstepcounter{subsubsection}
\noindent\textit{\thesubsubsection\quad Dataset licenses.}
\label{sec:case-model-dataset-licensing}

\subsubsubsection{Unlicensed datasets} %
Of the 96 observed and mappable datasets used by the top models, twelve (12\%) did not explicitly declare any licensing information in their tags. However, six of these datasets did declare at least some licensing information in other ways.  For example, \model{flax-sentence-embeddings/stackexchange\_xml} points users to an \textit{arxiv.org} link that includes more specific licensing information for the dataset, including steps for attribution. Two datasets (\model{kresnik/ zeroth\_korean} and \model{timdettmers/openassistant-guanaco}) %
 included licensing details in the text of their READMEs.  Another dataset, \model{openchat/openchat\_sharegpt4\_dataset}, provided a link to a GitHub repository that did include licensing information. However, the license in the repository is only explicitly stated to cover the codebase, not the dataset, though this may be implied.  Lastly, \model{cerebras/SlimPajama-627B} and \model{togethercomputer/RedPajama-Data-1T} were aggregated datasets that pulled from a variety of other datasets.  Their READMEs include links to the included datasets, leaving the user to determine the appropriate license based on which portions of the dataset are used. These examples highlight how despite the ability to automatically detect licenses in many cases, a human investigator may still need to perform manual analysis in others. Future work may investigate the usefulness of LLMs for this task.

\subsubsubsection{\other licenses}
Ten datasets declared that they were released under an \other license.  One of these was multi-licensed with other, more standard licenses, and in 5 of these cases the datasets were aggregates of other datasets or source code files under various licenses.  Hence, the \other label was being used to convey that the end user will need to research the specific data points or data subsets used, which could be a tedious and time consuming task.  Four datasets included licensing text/information in their READMEs.  These were not standard license texts and in some cases were very minimal.  For example, \model{dair-ai/emotion}  states only, ``The dataset should be used for educational and research purposes only.''  Another dataset, \model{jondurbin/airoboros-2.2.1}, provides text akin to a legal disclaimer: ``Much (most) of the data was generated via gpt-4 API calls, which has a restriction in the ToS about `competing' models. Please seek legal advice if you plan to build or use a model that includes this dataset in a commercial setting.''  The final dataset, which was multi-licensed, declares Open Portion of the American National Corpus, which is a collection of texts and not a license. It is possible this was meant to indicate that the license(s) of this corpus apply to the dataset, or at least portions of it.%

\subsubsubsection{The \unknown license}
Of the 96 observed and mappable datasets used by the top models, 8 (8\%) were released under the \unknown license tag. %
Two of these were dataset aggregates where the different data points were made available under different licenses, such that choosing a single license for the whole dataset would be difficult, if not impossible.  (The same situation that led other dataset owners to choose the \other license.) %
Three datasets had incomplete dataset cards/READMEs and did not include any licensing information.  One model reiterated that the licensing of the released dataset was \unknown but included no elaboration.  Two datasets stated that the licensing of the dataset depended on the licensing of its dependent data sources. One of these stated only that ``[t]he University of Washington does not own the copyright of the questions and documents included in TriviaQA.''  The other explained that ``[t]he licensing status of the dataset hinges on the legal status of the Pushshift.io data which is unclear.''  %
Despite using the \unknown license, presumably due to uncertainty regarding how their dataset should be licensed, %
six (75\%) datasets still included citation information. Thus, similar to the \other license, the \unknown license is not a monolith, encompassing a number of different scenarios. Here, we observed its use for aggregates with multiple licenses, insufficient licensing information, or on its own without further explanation.

\subsubsubsection{Multi-licensing}
\label{sec:case-model-datasets-multilicense}
There were two datasets that reported being released under multiple licenses.  The first of these, \model{nyu-mll/multi\_nli}, reported four licenses, including the \other license option.  Clarification concerning how these licenses relate is found in the associated dataset card, which explains that different portions of the dataset are released under different licenses.  The second example, \model{legacy-datasets/wikipedia}, reported two licenses, which also corresponded to portions of the dataset.  In both of these examples, we see that multi-licensing has been used to describe a complex licensing scenario with different subsets of a dataset.  This is contrary to the popular usage in open-source, where such a licensing scheme implies either an AND or an OR relationship between the licenses.

\subsubsubsection{License differences across models and datasets}
Given the difficulties in determining and interpreting the licensing for datasets that are multi-licensed or under the \other or \unknown license tags, we omit them from this analysis analysis. To better understand relationships between the licenses of models and datasets, we present an analysis considering the remaining 107 model-dataset license pairs. Twenty-four pairs had licensing information that exactly matched (\eg both the model and dataset were released under the \license{MIT} license).  (We note that a perfect match is not required, nor does it necessarily indicate full compliance with all licensing terms.)  Eighty-three pairs did not represent exact licensing matches.  For example, one model was unlicensed despite relying on seven datasets that had clearly defined licenses. \Cref{tab:sample_license_pairs} shows the most common pairings. Notably, the \license{cc-by-sa-4.0} and \license{cc-by-sa-3.0} licenses are restrictive and require derivative works to be released under the same license.  Depending on whether models are considered derivatives of their training data, the pairings (\license{apache-2.0}, \license{cc-by-sa-4.0}) and (\license{apache-2.0}, \license{cc-by-sa-3.0}) may represent a compliance issue since the share-alike term for the dataset licenses would have been violated. %

\begin{table}[]
\caption{Top 5 Model-Dataset license pairs in Top 187 models}
\label{tab:sample_license_pairs}
\begin{tabular}{lll}
\hline
\textbf{Model License} & \textbf{Dataset License} & \textbf{Count} \\ \hline
apache-2.0             & mit                      & 33             \\
apache-2.0             & cc-by-4.0                & 9              \\
apache-2.0             & cc0-1.0                  & 0              \\
apache-2.0             & odc-by                   & 8              \\
apache-2.0             & cc-by-sa-4.0             & 6 \\  \hline 
\end{tabular}
\end{table}

\section{Case study: top datasets}
\label{ssec:case_study_datasets}

Similarly to the above analysis of top models, we provide an analysis of the 100 most liked and 100 most downloaded datasets. 

\subsection{Sample construction} This sample of the top datasets on Hugging Face consists of 185 distinct datasets; fifteen datasets existed in both the top liked and the top downloaded. Across all 175,000 datasets in our full dataset, there are 49,466,329 total downloads. The 185 most popular datasets account for 41,994,158 of these, roughly 85\%.  This justifies that these models are worth further, more intensive investigation.

The datasets in the sample encompass 23 distinct tasks, including text generation (62), question-answering (35), fill-mask (13), text-classification (12), text2text-generation (9), multiple-choice (8), and summarization (7), and others with four or fewer occurrences. %

The sample contains 136 unique dataset owners. Only nine own more than two datasets in the sample, including Bigcode~(8), AllenAI~(5), EleutherAI~(5), Mteb~(5), Lighteval~(4), HuggingFaceFW~(3), HuggingFaceH4~(3), The Mozilla Foundation~(3), and BAAI~(3). %

Compared with the model analysis discussed previously, we notice that the distribution of datasets seems to be more decentralized.  Whereas the top 15 model uploaders owned 48.6\% of the top models, the top 15 dataset uploaders only owned 27.6\% of the top datasets. %

\subsection{Supply chain}

We observe a number of different dependency relationships among datasets in the sample. %
One such relationship type involves transforming the data in one dataset into another format, such as the \model{Cohere/wikipedia-2023-11-embed-multilingual-v3} dataset, which is a chunked and embedded version of the \model{wikimedia/wikipedia} dataset. We also observe aggregates of full datasets, such as \model{garage-bAInd/Open-Platypus} and \model{cerebras/SlimPajama-627B}, or aggregates of such aggregates, such as \model{teknium/OpenHermes-2.5}, which combines \model{garage-bAInd/Open-Platypus} (itself an aggregate) with various other datasets.

\subsection{Licensing}
\label{sec:case-dataset-licensing}

\subsubsection{Presence of licensing information}
To determine the presence of licensing for the datasets in our collection, we used a combination of automated and manual processes. \add{(We did not attempt to perform a legal analysis of any licensing decision, including whether any particular dataset was protectable under copyright law.)} First, we used the Hugging Face API to obtain licensing information from the dataset tags if it was available.  For any datasets where we were unable to identify a license through this process, we engaged in a manual analysis of the dataset card. Forty-four datasets required such analysis; of these, 9 (20\%) provided an exact license, while 14 (32\%) offered some type of related licensing information, such as rules governing how individual portions of the dataset were licensed. 
Licensing information sometimes had to be inferred from other sources; for example, the license of one dataset (2\%) could only be inferred from a linked GitHub repository, and 5 datasets (11\%) explained that they were aggregates of other datasets, encouraging users to ``Please refer to the licenses of the data subsets you use.'' We were unable to access one dataset, which has presumably been removed from Hugging Face or made private, and we were unable to locate any licensing information for the remaining 28 datasets.
Thus, in total, we obtained licensing information for 156 datasets (84\%) 
in our collection, though this information was not always sufficient to derive the exact licensing terms associated with the usage of a dataset. We note that this is lower than the 91\% observed in the most popular models. Several possible explanations exist for the lack of licensing information in datasets. For example, it might be the case that reconciling the licensing and copyrights of thousands of data points or even dozens of other datasets is outside the scope and expertise of most data scientists and model trainers. It might also be the case that datasets are often created with the end goal of training an AI model, and licensing/documenting them is not viewed as a priority by their creators. %

\subsubsection{Common licenses}
Here, we look at the 131 datasets %
that had identifiable, non-\unknown licenses %
(including custom licenses). %
Thirty-six (27\%) were released under \license{MIT}, 26 (20\%) under \license{Apache-2.0}, 11 (8\%) under \license{cc-by-4.0}, 10 (8\%) under \license{odc-by}, 9 (7\%) under \license{cc-by-sa-4.0}, 8 (6\%) under \license{cc0-1.0}, and 7 (5\%) under \license{cc-by-nc-4.0}.  We observe that, for the most part, these numbers mirror those for the entire ecosystem (as seen in \Cref{tab:top_licenses_datasets}), with the primary exceptions being the omission of the \license{openrail} license and the inclusion on the \license{odc-by} license.  We also note that a majority of the top datasets are permissively licensed.

\subsubsection{Analysis of the \unknown license}
\label{sec:case-study:unknown-license}
Six datasets were tagged with the \unknown license, for various reasons. One, \model{BAAI/COIG-PC}, was a dataset aggregate that provided rules on determining which licensing terms applied to which portions of the dataset. An additional three included no clear licensing guidance, and in fact, one hinted at violating the Terms of Service of a platform to create the dataset.  The license for one dataset could be inferred from a linked GitHub repository, but the certainty is low. Lastly, the bookcorpus dataset, instead of offering simple, actionable usage limitations or requirements, instead included a long essay on the complications and ethics associated with mining data from a site (in this case smashwords) where the data authors are unaware that their material could be used for training machine learning models and have therefore not given consent for that usage.  The dataset is still available on Hugging Face as of this writing, but the legality surrounding its use seems unclear even to the dataset owner.

\subsubsection{Analysis of the \other license}
\label{sec:case-study:other-license}
The sample also includes 16 datasets tagged with the \other license.  Of these, nine (56\%) were aggregates that had individual datasets or data points licensed under differing terms. We discuss the implications of these composite datasets in \Cref{sec:data-aggregates}. %
Four (25\%) were under a custom license or non-standard licensing text, such as a short statement about the intended use (\eg ``The dataset should be used for educational and research purposes only'') or conditional licenses (\eg ``If the source data of LIMA has a stricter license than CC BY-NC-SA, the LIMA dataset follows the same. Otherwise, it follows the CC BY-NC-SA license.''). %
The latter example illustrates how the task of license compliance---or, indeed, the task of determining the licenses within the component itself---is often left to the user. Two additional datasets (12\%) utilized the \other tag, but we were unable to locate any additional licensing information, rendering them effectively unlicensed. Lastly, we observed one dataset under the ``creative-commons-by-nc'' %
license. This nonstandard form, not present in Hugging Face's license list~\cite{hugging_face_licenses}, is likely intended to refer ``cc-by-nc-2.0,'' ``cc-by-nc-3.0,'' or ``cc-by-nc-4.0.'' Without additional information, however, it is impossible to determine with certainty the exact intended license, further highlighting the limitations of using plain text fields and motivating the need for validation.

\subsubsection{Analysis of Multi-licensed datasets}
\label{sec:case-study:multi-license}
Three datasets in the sample---\model{wikimedia/wikipedia}, \model{Salesforce/wikitext}, and \model{legacy-datasets/wikipedia}, all of which are comprised of text pulled from Wikipedia---reported more than one license in their tags, a combination of the \license{cc-by-sa-3.0} and \license{gfdl} licenses. We note that user-written text that is submitted to Wikipedia is generally under one of these two licenses as well. %
This example illustrates how multi-licensing schemes can attempt to capture the complexities of the licensing relationships of the individual data points that make up a dataset: both \model{wikimedia/wikipedia} and \model{legacy-datasets/wikipedia} provide nuance with statements such as, ``Some text may be available only under the Creative Commons license; see their Terms of Use for details. Text written by some authors may be released under additional licenses or into the public domain'' and ``Some text has been imported only under CC BY-SA and CC BY-SA-compatible license and cannot be reused under GFDL; such text will be identified on the page footer, in the page history, or on the discussion page of the article that utilizes the text.''  Lastly, although \model{Salesforce/wikitext} tags both licenses, in the body of its model card it simply states, ``The dataset is available under the Creative Commons Attribution-ShareAlike License (CC BY-SA 4.0),'' excluding any mention of the GFDL. In addition to these examples, there were three other datasets that provided no licensing information in their tags but were revealed to be multi-licensed through manual analysis: \model{uonlp/CulturaX}, \model{open-web-math/open-web-math}, and \model{Skywork/SkyPile-150B}.  All three explicitly stated that the datasets were released under the conditions of both licenses, representing an AND relationship (as described in \Cref{sec:multilicensing}). 
Recall that Hugging Face does not provide a built-in way to specify such a relationship, necessitating this clarification from the models. As shown from these examples, multi-licensing information can sometimes be obscured in a way that limits the abilities of users to understand which licenses apply in what manner, which can open the door to noncompliance by dependents.

\section{Discussion and Implications}
\label{sec:implications}
In this section we discuss the implications of our findings and potential directions for future work. %

\subsection{License compliance in the ML supply chain}
In \Cref{sec:rq1}, we showed that the structure of the ML supply chain on Hugging Face consists primarily of relatively short component chains. %
This presents a radically different landscape from the traditional software supply chain, in which a single piece of software can depend on potentially hundreds or even thousands of components. However, just because fewer dependencies are declared does not mean that ensuring license compliance is any easier in this ecosystem. In fact, as we observed, licensing information is often missing (\Cref{sec:licensing}) and there can be licensing differences between parent and child models in a model's lineage (\Cref{sec:base-to-deriv-variations}). Compounded with our observation that datasets may also have unclear licensing (\Cref{sec:common_licenses}) and may go undeclared by models (\Cref{sec:most-dependend-upon}), the state of documentation on Hugging Face appears to be counterproductive to successful license compliance. 
\looseness=-1

The task of managing different licenses across a model's components presents a novel challenge due to the advent of ML-focused licenses and a heavy reliance on \license{CC} licenses. 
While previous work has investigated compatibility between various traditional software licenses~\cite{kapitsaki2022towards,liu2024catch}, new licenses, such as the \license{OpenRAIL} family and the \license{Llama} licenses, require %
analysis to understand how they interact with existing licenses and with each other. Additionally, ML-focused licenses also indicate legitimate uses for the model and any derivatives, further separating them from the traditional open source definition. %
Proponents of \license{OpenRAIL} argue that existing open-source licenses are ``ill-adapted'' to handle ML components' differences from source code, particularly with respect to the responsible use of such components~\cite{openrailTowards}. %
This necessitates future work to provide a detailed analysis of software licenses with respect to ML components, including an analysis of these licenses and the ways in which licenses interact with ML components differently from traditional software.

It also becomes crucial to educate %
model owners and users on the terms associated with licenses.  For example, we observed that the simple naming requirements imposed by the \license{Llama3} license were followed in only 14.2\% of cases (\Cref{sec:naming_requirements}).  This suggests either extensive deliberate noncompliance or, more likely, ignorance of the relevant license terms. To concretely identify the cause and provide potential mitigations, future work should investigate developers' views and understanding of licenses in the ML supply chain.

It also remains unclear, particularly with the relationship between datasets and models, how exactly license terms, such as \license{GPL}'s copyleft, should be applied \cite{ossForML}.  To what extent should a model be considered a derivative work of the dataset(s) on which it was trained?  More concretely, if a model trainer trains a model on a dataset governed by a copyleft license and then distributes the model, but not the dataset, has the trainer violated the dataset license?  \add{More fundamentally, to what extent are ML datasets protected by copyright law in various jurisdictions? In other words, is a particular selection and arrangement of data protected by copyright, or does copyright attach only at the level of individual data, if at all?} These questions and many others will likely need to be resolved by legal teams, regulators, and possibly the courts.

We observed challenges presented by licensing datasets that contain data points under different licenses (\Cref{sec:case-study:unknown-license,sec:case-study:other-license}). While these challenges are not new \cite{longpre2023data,lee2023ai}, future research and the Hugging Face development team should investigate  more intuitive ways of conveying that the licensing of a dataset depends on the specific portions used rather than allowing users to obscure this information by the use of an \other label, instead reserving the \other label for custom or non-standard licenses. Relatedly, dataset and model repositories should encourage model/dataset uploaders to select from already existing licenses rather than using \other in order to avoid rampant license proliferation.  If custom license text is required, it should be within a LICENSE file so that it is easy for users to locate and access.

Lastly, in \Cref{sec:multilicensing,sec:case-study:multi-license,sec:case-model-datasets-multilicense}, we also observed instances of ambiguity introduced through multi-licensing.  Determining the intended relationship among the multiple licenses assigned by the model/dataset owner can be a difficult task, and different pitfalls come with the assumption of an OR versus an AND relationship.  Model hubs should include standardized ways for model/dataset owners to express these relationships and educate users on why they might want to use a multi-licensing scheme.  %

\subsection{Dataset aggregates}
\label{sec:data-aggregates}
As discussed in \Cref{sec:case-study:unknown-license,sec:case-study:other-license}, some datasets on Hugging Face were aggregates or composites of other datasets.  In these situations,  the uploader would leave license identification to the end user, saying something to the effect of ``Look at the licenses for the different subsets of this dataset.''  %
In addition to raising potential compliance questions, the lack of standardization in the way that this aggregate nature was conveyed potentially presents challenges.  We observed model uploaders using the \unknown and \other license options to refer to this composite nature.  Other uploaders opted to leave their dataset unlicensed in the tags and elaborated in the body of the dataset card that the terms of the individual licenses needed to be complied with.  Still others opted to declare multiple licenses in their tags, not to indicate an AND or an OR relationship but rather to indicate that different licenses applied to different subparts of the dataset (\Cref{sec:case-study:multi-license}).  Without an additional field to allow owners to describe this type of licensing situation more precisely, this structure is likely to contribute to developers' difficulty with understanding licensing tasks, a difficulty that previous work has explored~\cite{almeida2019investigating}.%

The purpose of these dataset aggregates would seem to be assembling large quantities of training data for machine learning models and making them easily accessible. %
Additional problems arise if the licenses contained within an aggregate conflict.  For example, \model{nguha/legalbench} is composed of datasets that are under two different copyleft CC licenses (\license{Attribution-ShareAlike 4.0 International} and \license{Attribution-NonCommercial-ShareAlike 4.0 International}).  It is not possible to meet the terms of both licenses, meaning that this dataset composite and models trained on this dataset risk being deemed noncompliant. This also poses risks for any end user, who might not have the motivation or the knowledge to determine compliance, raising nonfunctional ethical considerations that are deserving of attention by researchers.

\subsection{Potential difficulties in using model cards as ML/AIBOMs}
Amidst the growing push for better transparency and security in software through Software Bills of Materials (SBOMs)~\cite{xia2023empirical, stalnaker2024boms}, calls have been made for similar BOMs for ML components~\cite{stalnaker2024boms, xia2023empirical, barclay2019towards, barclay2022providing}. These documents serve as inventories of all components within a piece of software and can document a variety of the software's traits, including dependencies, licensing, and security. Distinct ML/AI Bills of Materials (ML/AIBOMs) may be necessary to address such components' different inputs, security concerns (such as model poisoning), and ethical considerations~\cite{xia2023empirical, xia2023trust, bi2024way, lu2024taxonomy}. The SPDX \cite{spdx} Working Group, operating under the auspices of the Linux Foundation, is developing guidelines and a proposed standard for ML/AIBOMs~\cite{aibom2024} that will complement the SPDX ISO Standard ISO/IEC 5962:2021, which describes the use of SBOMs to document the components used in creating a software system~\cite{ISO5962-2021}. %
CycloneDX \cite{cyclonedx} is also working on support for ML/AIBOMs~\cite{cyclonedx-mlbom}. %

Prior work has also suggested that model and data cards can serve as an ML/AIBOM~\cite{stalnaker2024boms}, or that the information they provide can assist with their creation~\cite{xia2023trust, pepe2024hugging}, but our work suggests that these tools are not yet robust enough to serve this purpose. As we highlight in \Cref{sec:rq0}, the information provided by model cards is often missing key elements, including datasets that the models were trained on. This immediately obviates many of the benefits of ML/AIBOMs, including understanding licensing obligations that might be associated with that dataset, being aware of potential model poisoning attacks, and providing the ability to select models trained on ethically sourced data. Additionally, instances in which model cards are locked behind Terms of Service agreements or other restrictions (\Cref{sec:no_api_access}) could further limit their usefulness as ML/AIBOMs to consumers. Our work provides additional evidence that, in practice, model cards contain little actionable information, making them difficult to use as ML/AIBOMs~\cite{bhat2023aspirations}.

\subsection{The need for follow up surveys and interviews}
Our study raises many questions concerning what motivates developers when they make decisions concerning documentation, model reuse, and licensing.  %
Conversations with Hugging Face developers through surveys and interviews can help to bridge this gap.  There are numerous key questions such work should consider. What factors influence developers when choosing a license for their dataset/model \add{and what do they believe the scope of that license to be} (\Cref{sec:common_licenses})? Why do developers so commonly opt to use an \other license (\Cref{sec:common_licenses})? What motivates developers to multi-license a model/dataset and what is the intention behind the license interaction (\Cref{sec:multilicensing})?  Why do so few developers fine-tuning the \model{llama3} model comply with the naming requirements imposed by the \license{llama3} license (\Cref{sec:naming_requirements})? What is the primary motivation for developers to write documentation for their models?  For each of the various stakeholders in the space (model owners, fine-tuners, and users), which metadata fields do they perceive as most important?  Why are some models listed as datasets (\Cref{sec:models_as_data})? Answers to these questions will inform how future tools, practices, techniques, and even educational materials should be developed to have maximal impact on the emerging ML community. %

\subsection{The addition of fields for recording model relationship information}

As the relationships between ML models become more complex, it becomes increasingly important to keep a record of those relationships (fine-tuned, quantized, ensemble, distillation, \etc).  This information is crucial not only for transparency and understanding how the model was created but also for license compliance.  Just as in traditional software, licenses may have terms that govern the manner in which the component may be (re)used and distributed. For example, many generative models disallow using generated content to train a derivative model~\cite{openai_terms}. As noted in \Cref{sec:no_relationship}, Hugging Face provides no standardized way to express relationship information.  This inability to specify relationships may have resulted in some model owners declaring models used in student-teacher relationships as datasets (\Cref{sec:models_as_data}).  We suggest that model hubs provide a standardized, machine-readable way of conveying this type of information.  Just as Hugging Face currently provides a standard list of architectures, it could also provide a curated list of relationship types.  
\looseness=-1

\subsection{The need to use unique identifiers}
While unique identifiers already exist on Hugging Face, they are not currently being used by model/dataset owners.  This can result in problems, particularly when trying to trace model/dataset provenance.  Model hubs should encourage their users to use these existing unique identifiers, perhaps by including tools that facilitate the creation of model cards or the provision of metadata.  For example, a developer could supply a colloquial name for a model hosted on the platform, such as the familiar owner/model convention, and the tool could automatically conduct a search and, where possible, map the provided name to the unique ID.  This would immediately alert the developer to typos, remove ambiguities, and prevent the name drift issues we discussed in \Cref{sec:naming-problems,sec:missing_refs}. Furthermore, such a system could potentially allow for automatic updates (or request manual developer updates) when the information for downstream elements changes, thereby facilitating an ecosystem which is more supportive of supply chain management tasks and lifting the burden from downstream developers.

\subsection{Differences between the ML supply chain and traditional software supply chain}
\label{sec:supplychaindifferences}
Our findings highlight a notable difference between the ML supply chain and the traditional software supply chain.
In many software ecosystems with dependency management, dependency information for components in the software supply chain is organized in manifest files such as \texttt{pom.xml}, \texttt{package.json}, or \texttt{requirements.txt}.
These files can then be used to generate other dependency tracking documents like SBOMs and have also been used by forges like GitHub to build dependency graphs \cite{dep_graph}.  Notably, however, these manifest files have utility %
beyond dependency tracking.  They are necessary for setting up a fresh installation of the software---that is, the list of dependencies \add{that} must be downloaded in order for the software to function properly.  This is not the case with the information supplied by developers in model cards.  Because erroneous mappings to datasets and base models do not leave the model in question unusable as a practical matter, it is therefore easier for typos and other mistakes to go unnoticed and thus uncorrected.
This disconnect between correctness and functionality necessitates tooling that has similar functionality and richness to dependency management tools for traditional software, as it leaves ample opportunity for errors without comparable means of checking for and correcting them.

\subsection{Quality control for model hubs}
Our analysis of the model documentation available on Hugging Face shows that the information that developers provide is often incomplete, malformed, or available only in human-readable text. We observed some documentation issues, such as the presence of duplicates in lists of datasets and base models (\Cref{sec:no_relationship}) or the declaration of a model as its own base (\Cref{sec:missing_refs}) that could be easily corrected by automated tools like linters. Additionally, an automated tool could also provide standardization for data fields to facilitate metadata analysis.  For example, the variants for how one can specify that there are no datasets or base models ([], ``'', and \textsc{None}) should be standardized to a single type.  We are not suggesting that model hubs should be responsible for curating the metadata of all models on their platforms, but where automated checks can efficiently be put into place and this is easy to implement, they should be employed. 

Potential solutions could closely resemble approaches to automatically check the quality of bug reports. Similarly to how tools can check for important components of a bug report (\eg expected behavior, steps to reproduce)~\cite{chaparro2017detecting}, in theory, tools may be able to check for missing information in model cards. %
Some work is already emerging in this area, such as DocML \cite{bhat2023aspirations} and AIMML \cite{TsayBHSM20}; however, more work needs to be done to understand what suggestions should be made to model owners, how to handle complex licensing scenarios, and how to best implement tools that assess ML documentation. %

Since license metadata for models and datasets is entered manually, there is the possibility that typos are introduced.  We encountered instances where the standard SPDX identifier \license{cc-by-sa-4.0} was mistakenly entered as \license{cc-by-sa 4.0} (\ie missing the final hyphen).  While not a major issue, typos like these necessitate that tools perform data cleaning in addition to any work related to license compliance.  Again, it would be preferable if Hugging Face provided a basic linter/validator for model/dataset cards and provided users with a drop-down/auto-complete from which to select popular licensing options, if this would be easy to implement. %

\subsection{Potential pitfalls when mining Hugging Face}
\label{sec:pitfalls}
The numerous documentation challenges that we discuss in \Cref{sec:rq0} demonstrate that researchers relying on Hugging Face mining to collect information on the ML supply chain must proceed thoughtfully, with caution and care, to avoid potential pitfalls. %
Unlike source code forges like GitHub, it seems that some model owners may be uploading and publishing models as a by-product of their workflows \cite{model_uploading,model_sharing}. %
As such, we speculate that some models are made publicly available not because the owner intends them to be reused by the community but rather because the owner is using the resources of Hugging Face, Google Colab, or some other cloud provider~\cite{google_cloud_collab}. %
This might explain why we observed so many models with the same names, such as \model{my\_awesome\_model} which is the example name used in the documentation for uploading a PyTorch model~\cite{model_uploading}.  (See \Cref{sec:naming-problems}.)  More research is required to understand how such tools might populate the supply chain with such ``intermediate'' models or those that are otherwise not intended for public consumption. To mitigate the risk such models pose, future studies should be mindful of the models they include in their analysis, erring toward those that are more popular.  %
However, we caution researchers to be thoughtful in how they determine model popularity.  In the context of the supply chain, we believe that downloads is the more appropriate metric, as opposed to likes. As shown in \Cref{tab:top_datasets_likes}, the most liked ``dataset'' on Hugging Face is in reality a repository curating ChatGPT system prompts (\eg ``Imagine you are an experience Ethereum developer...'').  Additionally, the repository for \model{lllyasviel/sd\_control\_collection} (with 1,574 likes, but 0 downloads) is actually a collection of various models assembled ``for users to download flexibly.''  Future studies will be left to explore the prevalence of curated repositories hosted as models/datasets and the exact nature of potentially ad hoc developer workflows.
\looseness=-1

\subsection{Recommendations for Stakeholders}
Here, we offer concise and actionable recommendations for various stakeholders involved in the ML supply chain based on our results.%

\subsubsection{Model trainers and dataset curators}
\begin{enumerate}
    \item When licensing models and datasets, use tags, the \cardData attribute, and/or a LICENSE file rather than including licensing information in the README or via an external source.
    \item When releasing a model or dataset under a multi-licensing scheme, clearly indicate how the licenses relate.
    \item When choosing a license for models and datasets, where possible use common, pre-existing licenses, rather than drafting a new license or modifying the terms of an existing one.
    \item When there is no choice but to draft a new license, make sure that the full license text is made available in a LICENSE file available on the model/dataset repository.  Also ensure that licensing terms are unambiguous and comprehensive (\eg more than ``This model is open source''). 
    \item When using models and datasets, particularly for larger or commercial projects, verify that the declared license is consistent and compliant with other upstream sources.
\end{enumerate}

\subsubsection{Model and dataset sharing platforms}%
\begin{enumerate}
    \item Provide intuitive ways for dataset curators to specify the licenses of their work, particularly in the case of dataset aggregates.  Simply using the \other or \unknown license options or declaring multiple licenses without clarification seems insufficient. %
    \item When designing documentation tools for model trainers and dataset curators, offer user-friendly options, such as multi-select widgets ( with an \other option) and smart auto-complete text fields, instead of simple plain text fields.  This is particularly the case for for fields that have a finite set of recognizable values, such as licenses, pipelines, \etc  %
    \item Perform basic validation of model/dataset metadata, including checking for obvious typos, needless repetitions, or improper uses of fields.
    \item Automatically include unique identifiers for models and datasets listed as dependencies when possible.  Human readable names for both models and datasets can change over time, but humans are also not well suited for working with the random strings of characters often associated with unique identifiers. When models/datasets are referenced in the metadata of other models/datasets, they should also be accompanied with a unique identifier that anchors the dependency, even if the human readable name should change. For Hugging Face, this might mean automatically adding the internal model ID as part of the base\_model information in the \cardData attribute.%
    \item Offer standardized, precise ways for model trainers to specify the dependency relationships between models (fine-tune, distillation, quantization, \etc), as these may have implications for various supply chain management activities, including licensing compliance.
    \item For platforms that host both models and datasets, allow model trainers to specify where datasets are located with an external link if they are not available directly on the platform in order to make various provenance tasks significantly easier.
\end{enumerate}

\subsubsection{SE and Law Researchers}
\begin{enumerate}
    \item Be aware and cautious of the limitations and quirks of mining data from Hugging Face.  For example, the readily available ``downloads'' metadata field actually only reports downloads that have occurred within the last 30 days.  %
    Total downloads can be accessed on a per query basis by providing additional arguments through the API, but it is not possible (as far as the authors can tell) to have the API return a list of models ordered by total downloads.
    \item Choose proxy metrics carefully.  As described in \Cref{sec:pitfalls}, when using model/dataset popularity/engagement as a proxy for quality or importance, be mindful that like count (similar to GitHub's stars) may not be as reliable as the download count.
    \item Understand that dependency information in the ML supply chain is more difficult to validate than dependency information in the traditional supply chain as described in \Cref{sec:supplychaindifferences}.
    \item Develop updated guidance and compliance matrices that can support stakeholders in the ML ecosystem as they work with various different licenses in a novel context.
    \item Investigate developers’ views and understanding of licenses in the ML supply chain through surveys and interviews.
    \item Know that not all models on Hugging Face were uploaded with the intent to publish. Instead, some may be incidental uploads that are a by-product of developer workflows through which they access cloud computing resources.
\end{enumerate}

\section{Threats to Validity}
\label{sec:threats}

\textbf{Construct Validity.} 
We rely on information that is available through the Hugging Face API in a machine-readable format. We addressed cases where information was missing from this source by applying Python's \texttt{request} and BeautifulSoup libraries to retrieve missing information. However, this means that we likely missed licensing and dependency information that was presented in natural language.  We measure and mitigate this risk with a review of the top 100 ``unlicensed'' models sorted by number of downloads, which showed that the majority did not, in fact, declare a license or declared a license in a non-standard or transitive way.  %
 
\textbf{Internal Validity}.
We followed best practices in our mining and data cleaning approaches. Conclusions were drawn based on empirical evidence found in the data. We relied on tags applied to components to detect their licenses, but such information might be present in other, human-readable forms or through links to other resources, such as a research paper or another repository. 
The labeling of license categories is based on our understanding of the licenses. To mitigate bias in these labels, two researchers examined each license to confirm the label applied to it. %

\textbf{External Validity.}
The generalizability of our findings is restricted by our choice of Hugging Face, which might not be fully generalizable to all model-sharing platforms. It is possible that we missed other challenges and obstacles faced by other model hubs or that different model-sharing sites have different levels of documentation and dependency management techniques, given our assumption that highly-downloaded models tend towards better documentation. We also observed that relatively few models and datasets declared licensing information, and models often did not declare any datasets. This implies that a great portion of the ecosystem may yet be undocumented. As such, our results may be biased towards the components that do provide documentation, and may be less generalizable to those which do not. We attempt to mitigate this risk by providing a detailed analysis of top-rated components to capture the specific challenges facing components that are used frequently and are thus applicable to many users. Additionally, given that the field of AI is rapidly evolving and new models are created every day, the conclusions we reached might be different if we were to take an updated snapshot of the supply chain at some future point. The extent to which this factor might change our results is unknown, but we maintain that it is important to capture an understanding of this nascent supply chain so that future investigations can understand the ways in which the supply chain has changed over time.

\section{Conclusion}
In this work we conduct a comprehensive mining study and analysis of the ML supply chain on Hugging Face. We construct a reusable dataset of over 750,000 models and 175,000 datasets, which we make freely available in our replication package \cite{anonymous_repo}.  Using this data we create a supply chain graph that illustrates the relationships between ML models and the models they depend on, describe the notable features and properties of this supply chain, analyze the current state of documentation on Hugging Face, and consider the current state of licensing and license compliance in the ecosystem.  We offer a rich discussion of the challenges that not just Hugging Face but any sizable model hub will encounter.  Lastly, we offer concrete and actionable recommendations for various stakeholders involved with the ML supply chain that can be adopted to facilitate better supply chain management.

\section*{Data availability}
We provide a replication package %
containing the raw mined data, our cleaned and normalized dataset, scripts for data cleaning, mining utilities, analysis scripts, and other data required for verifiability~\cite{anonymous_repo}. %

\section{Acknowledgments}
This research has been supported in part by NSF grant CCF-2217733. Any opinions, findings, and conclusions expressed herein are the authors’ and do not necessarily reflect those of the sponsors. A complete, detailed list of image attributions can be found at \cite{anonymous_repo}.

\bibliographystyle{ACM-Reference-Format}
\bibliography{references}

\end{document}